\documentclass[superscriptaddress,aps,letterpaper,10pt,twocolumn,floats,amsmath,amsfonts,amssymb,prb]{revtex4-2}

\usepackage{graphicx}
\usepackage[colorlinks=true,citecolor=blue]{hyperref}
\usepackage[caption=false]{subfig}
\usepackage{pstricks}
\usepackage{pst-node}
\usepackage{amsmath}
\usepackage{amsbsy}
\usepackage{verbatim}
\usepackage{float}
\usepackage{amsthm}
\usepackage{dsfont}

\newtheorem{definition}{Definition}

\begin{document}
	
	
\title{System-bath entanglement of noninteracting fermionic impurities: \\
Equilibrium, transient, and steady-state regimes}

\author{Krzysztof Ptaszy\'{n}ski}
\email{krzysztof.ptaszynski@uni.lu}
\affiliation{Complex Systems and Statistical Mechanics, Department of Physics and Materials Science, University of Luxembourg, L-1511 Luxembourg, Luxembourg}
\affiliation{Institute of Molecular Physics, Polish Academy of Sciences, Mariana Smoluchowskiego 17, 60-179 Pozna\'{n}, Poland}

\author{Massimiliano Esposito}
\affiliation{Complex Systems and Statistical Mechanics, Department of Physics and Materials Science, University of Luxembourg, L-1511 Luxembourg, Luxembourg}

\date{\today}

\begin{abstract}
We investigate the behavior of entanglement between a single fermionic level and a fermionic bath in three distinct thermodynamic regimes. First, in thermal equilibrium, we analyze the dependence of entanglement on the considered statistical ensemble: for the grand canonical state, it is generated only for a sufficiently strong system-bath coupling, whereas it is present for arbitrarily weak couplings for the canonical state with a fixed particle number. The threshold coupling strength, at which entanglement appears, is shown to strongly depend on the bath bandwidth. Second, we consider the relaxation to equilibrium. In this case a transient entanglement in a certain time interval can be observed even in the weak-coupling regime, when the reduced dynamics and thermodynamics of the system can be well described by an effectively classical and Markovian master equation for the state populations. At strong coupling strengths, entanglement is preserved for long times and converges to its equilibrium value. Finally, in voltage-driven junctions, a steady-state entanglement is generated for arbitrarily weak system-bath couplings at a certain threshold voltage. It is enhanced in the strong-coupling regime, and it is reduced by either the particle-hole or the tunnel coupling asymmetry.
\end{abstract}

\maketitle

\section{Introduction}
The notion of entanglement refers to genuine quantum correlations between two or more physical objects that cannot be explained by any classical model~\cite{horodecki2009}. In addition to its fundamental importance and applicability as one of the basic resources in quantum technology, entanglement has attracted attention in the field of condensed matter physics, as it provides important information on the behavior of quantum many-body systems~\cite{amico2008, lafrorencie2016, chiara2018}. In the context of open quantum systems, consisting of a system attached to one or more thermal baths, most studies focused on the issue of how entanglement between two constituents of the system is affected by the interaction with the bath. Among others, these investigations dealt with the relation between entanglement decay and (non-)Markovianity of the system~\cite{bellomo2007,mazzola2009,maniscalo2008,bellomo2008,rivas2010}, as well as entanglement generation through system-bath interaction in both transient~\cite{plenio1999, kim2002, jakobczyk2002, braun2002, benatti2003} and steady-state~\cite{schneider2002, kraus2008, kastoryano2011, krauter2011, barreiro2011, lin2013, shankar2013, brack2015} regimes. Much less of the studies were concerned with the entanglement between the system and the bath. This is understandable as the dimension of the Hilbert space of the bath increases exponentially with its size. As a consequence, characterization of the system-bath entanglement -- which requires knowledge of the total system-bath state -- is usually very difficult. 

Previous studies of entanglement between the system and the bath can be divided into two groups. The first focused on static properties of entanglement in the ground or thermal state of the total system-bath Hamiltonian. Among others, such entanglement has been used to shed light on the paradigmatic model of strongly correlated physics, namely the Kondo model~\cite{kondo1964}. In particular, it has been applied to investigate the finite-temperature behavior of the Kondo effect~\cite{lee2015, shim2018, kim2021}, spatial extent of the Kondo cloud~\cite{bayat2010, lee2015, wagner2018, shim2023}, competition between screening channels in the multichannel Kondo effect~\cite{alkurtass2016, bayat2017, kim2021}, or the quantum critical behavior in the two-impurity Kondo model~\cite{bayat2012, bayat2014, bayat2017}. Furthermore, certain studies demonstrated the connection between entanglement and observable quantities, such as electric conductance~\cite{yoo2018} or thermometric sensitivity~\cite{mihailescu2022}. Entanglement was investigated also for the spin-boson model~\cite{costi2003, kopp2007, lehur2007, lehur2008}, but, to our knowledge, only in the zero-temperature case. These studies revealed, e.g., a nonanalytic behavior of the entanglement entropy at the quantum phase transition point~\cite{lehur2007}.

The second group of studies analyzed the dynamic properties of the system-bath entanglement for a system initialized out-of-equilibrium with respect to the bath. In particular, they mainly focused on the case of pure dephasing (i.e., dynamics which does not change the state populations in a specified basis). For such a situation, Roszak~\cite{roszak2018} provided analytic criteria for the presence of entanglement in a generic open quantum system. More specifically, entanglement has been found to be not necessary for decoherence~\cite{maziero2010, roszak2015}, but crucial for the emergence of classical objectivity within the framework of quantum Darwinism~\cite{roszak2019, garciaperez2020, megier2022}.  Beyond the paradigm of pure dephasing, Eisert and Plenio~\cite{eisert2002} investigated entanglement in the quantum Brownian motion model (i.e., Caldeira-Leggett model~\cite{caldeira1983}). It was shown that entanglement is always immediately generated for a pure initial state of the system; on the other hand, for any system-bath coupling there exists an initial mixed state of the system and the temperature of the bath for which entanglement is absent at all times.

Furthermore, a few studies investigated possible connections between the system-bath entanglement and the strong-coupling thermodynamic effects. First, Refs.~\cite{allahverdyan2000, nieuwenhuizen2002, allahverdyan2002, horhammer2005} observed a so-called violation of Clausius inequality $\Delta S_B=-\beta Q$ [where $S_B=-\text{Tr} (\rho_B \ln \rho_B)$ is the von Neumann entropy of the bath and $Q$ is the heat extracted from the bath], and related it to the system-bath entanglement. This assertion was later questioned by Hilt and Lutz~\cite{hilt2009}, who showed that the relation $\Delta S_B=-\beta Q$ can be violated also for separable states. As further discussed in Ref.~\cite{ptaszynski2019}, this violation is rather common in nonequilibrium settings and is not even restricted to the strong-coupling regime. Recently, Ref.~\cite{bernardo2021} observed a proportionality between the system-bath entanglement and interaction energy for a bath consisting of a single qubit, reaching the conclusion that the imbalance between the energy changes of the system and the bath is responsible for the generation of entanglement.

\begin{figure}
	\centering
	\includegraphics[width=0.8\linewidth]{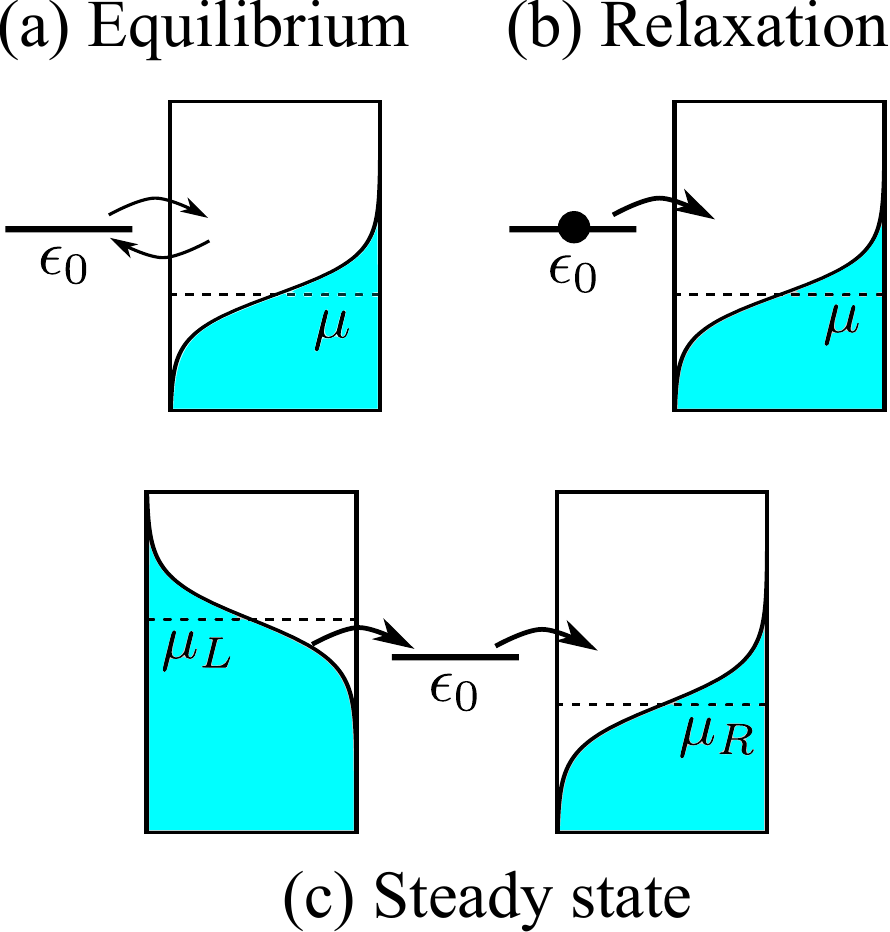}
	\caption{Schematic presentation of thermodynamic regimes considered in the paper: (a) equilibrium state of the impurity and the bath, (b) transient relaxation of the impurity to equilibrium, (c) steady-state transport between two reservoirs with different chemical potentials.}
	\label{fig:schem}
\end{figure}

In this paper we study the system-bath entanglement in one of the paradigmatic models of open quantum systems, namely, the noninteracting resonant level. It consists of a single fermionic level coupled to a noninteracting fermionic bath through bilinear tunneling Hamiltonian. We will focus on three distinct thermodynamic regimes, schematically presented in Fig.~\ref{fig:schem}: (a) global thermal state of the system and the bath, (b) transient relaxation of the system initialized in an out-of-equilibrium state, and (c) steady-state transport between two reservoirs driven by the applied voltage. Our motivation is, on the one hand, system-specific, aiming to investigate specific features of fermionic entanglement. On the other hand, as we will show, our study leads to more general insights into the relation between entanglement and (non-)Markovianity, strong-coupling thermodynamic effects, and nonequilibrium driving. The goals of our paper are described in more detail below.

\textit{Thermal equilibrium}.---As previously mentioned, much attention has been devoted to the thermal entanglement in the Kondo model of a spin coupled to a fermionic bath via the exchange interaction. This model further corresponds to the low energy regime of the Anderson model, namely, a Coulomb-interacting fermionic impurity tunnel-coupled to a fermionic bath~\cite{anderson1961, schrieffer1966}. It is then natural to ask which features of entanglement in the Kondo and Anderson models can be directly related to strong correlations, and which can be already observed in the noninteracting case of a vanishing Coulomb coupling. Surprisingly, this topic has so far only been scarcely studied in the literature. To the best of our knowledge, the role of interelectron interaction strength has been investigated only in Ref.~\cite{shim2018}, with the analysis of the noninteracting case restricted to the zero-temperature regime. At the same time, a detailed analysis of the conditions of the generation of equilibrium entanglement has been performed for an analogous noninteracting bosonic model of a single quantum harmonic oscillator coupled to a harmonic bath (the Caldeira-Leggett model~\cite{caldeira1983}), which revealed a sudden death of entanglement above a certain threshold temperature (dependent on the coupling strength to the bath)~\cite{hilt2009}. Our study aims to fill this gap by analyzing the dependence of system-bath entanglement on different system parameters, as well as the dependence on the considered thermodynamic ensemble.

\textit{Transient dynamics}.---The interaction of an open quantum system with the bath leads to the generation of system-bath correlations. At first glance, one might expect that such correlations are negligible in the validity regime of the Born-Markov approximation (used to derive the Markovian master equation for the reduced dynamics of the system), which assumes that at all times the global state of the system $S$ and the bath $B$ can be approximated as a product state $\rho_{SB}(t) \approx \rho_S(t) \otimes \rho_B^\text{eq}$, where $\rho_B^\text{eq}$ is the thermal equilibrium state of the bath. Indeed, certain types of system-bath correlations can be directly related to the non-Markovianity of the reduced dynamics~\cite{mazzola2012, smirne2013, campbell2019, strasberg2019}. However, it has been shown that effective Markovianity of dynamics does not necessarily imply the absence of correlations defined in information-theoretic terms, such as the quantum mutual information (at least at short timescales)~\cite{ptaszynski2022, pucci2013, einsiedler2020, colla2021, ptaszynski2023}. Rather, as shown for collisional models, only a part of the system-bath correlations is relevant for the reduced dynamics~\cite{campbell2018}. Furthermore, different microscopic models can generate the same reduced dynamics but different system-bath correlations~\cite{smirne2021}, further demonstrating the lack of an obvious link between them.

The question then arises whether the character of the reduced dynamics is related to the behavior of genuine quantum correlations, such as entanglement. It has sometimes been argued that the validity of Born-Markov approximation precludes the presence of entanglement~\cite{shresta2005, kosloff2013}; however, as discussed in the previous paragraph, such a relation is not necessarily obvious for correlations defined in information-theoretic terms. Indeed, for the case of pure dephasing, Refs.~\cite{pernice2011, pernice2012} found no obvious link between the system-bath correlations and non-Markovianity: entanglement may appear also during Markovian dephasing, though often on timescales longer than the decoherence time. Similarly, Ref.~\cite{szankowski2020} found no connection between the presence or absence of entanglement and the possibility of describing the dephasing using a classical noise. In this study, our aim is to explore this issue in the context of relaxation dynamics of a fermionic impurity.

We are also interested in the relation between entanglement and thermodynamics. First, we are motivated by the observation that for a properly thermalizing bath (which is determined, e.g., by the density of states in the bath) the state of the system relaxes over time to equilibrium corresponding to the global Gibbs state of the total system-bath Hamiltonian~\cite{trushechkin2022}. We want to investigate whether and when this is also true for the system-bath entanglement. Second, we want to verify the alleged link between the transient entanglement and the interaction energy~\cite{bernardo2021}.

\textit{Steady state}.---Finally, while entanglement within nonequilibrium steady states of many-body systems has already received a certain attention~\cite{gullans2019, panda2020, fraenkel2022, eisler2022, fraenkel2023}, to the best of our knowledge no study focused specifically on entanglement between a small impurity and the bath. Instead, previous investigations of fermionic~\cite{sharma2015, dey2020} and bosonic~\cite{sable2018} impurities considered quantum mutual information, which does not distinguish between classical and quantum correlations. These works demonstrated an increase of the steady-state value of system-bath correlations with the applied voltage or temperature bias. As the nonequilibrium driving of open quantum systems may lead to generation of steady-state intrasystem entanglement, we aim to explore whether this conclusion can be generalized to the system-bath entanglement~\cite{schneider2002, kraus2008, kastoryano2011, krauter2011, barreiro2011, lin2013, shankar2013, brack2015}.

\textit{Structure of the paper}.---This article is organized as follows. In Sec.~\ref{sec:ferment} we present the definition of entanglement applicable to fermionic systems and methods used to quantify the system-bath entanglement. In Sec.~\ref{sec:model} we discuss the  model and the methods used to describe its dynamics. Secs.~\ref{sec:equi}--\ref{sec:volt} present the results for the equilibrium, transient relaxation, and voltage-driven cases, respectively. Finally, Sec.~\ref{sec:concl} brings the conclusions following from our results. Appendices~\ref{sec:transp}--\ref{sec:mutinf} contain the definition of partial transposition, the description of the Householder tridiagonalization algorithm, and the analytic theory of system-bath mutual information.

\section{Fermionic entanglement} \label{sec:ferment}
\subsection{Entanglement definition}
Let us first discuss how we define the system-bath entanglement for fermionic systems and how its presence can be detected. A standard definition of entanglement used in quantum information theory states that the bipartite system $SB$ is deemed entangled when it is not separable, i.e., when its density matrix $\rho_{SB}$ cannot be written as a classical mixture of tensor product states~\cite{nielsen2010}, 
\begin{align} \label{statebased}
	\rho_{SB} =\sum_{\nu} p_\nu \rho_{S}^\nu \otimes \rho_{B}^\nu,
\end{align}
where $p_\nu$ are positive-valued probabilities summing up to 1 and $\rho_{S}^\nu$, $\rho_{B}^\nu$ are positive semidefinite matrices with trace 1. For fermionic systems, however, the proper definition of entanglement is a more subtle issue due to the parity superselection rule, which prohibits coherent superpositions of states with even and odd particle parity~\cite{wick1952, szalay2021}. This rule provides constraints on the physically allowed states, observables, and operations~\cite{vidal2021}. As thoroughly discussed by Ba\~{n}uls \textit{et al.}~\cite{banuls2007}, applying the parity superselection rule in different ways, one obtains a hierarchy of definitions of entanglement, which may be either weaker~\cite{dariano2014,dariano2014b} or stronger~\cite{wiseman2003, ding2021, ding2022, ernst2022} than the standard one. In this paper, we use the most stringent notion, previously applied in Refs.~\cite{wiseman2003, ding2021, ding2022, ernst2022, ding2024}:
\begin{definition}[Observable-based definition of fermionic entanglement] \label{obsbased}
	Let us first define the locally projected state
	\begin{align}
		\pi_{SB}=\sum_{\alpha,\gamma=e,o} \left( \mathbb{P}_\alpha^S \otimes \mathbb{P}_\gamma^B \right) \rho_{SB} \left( \mathbb{P}_\alpha^S \otimes \mathbb{P}_\gamma^B \right),
	\end{align}
	where $\mathbb{P}_{e/o}^X$ is the local projection of the subsystem $X \in \{S,B\}$ on the even/odd particle parity sector. Then, the state $\rho_{SB}$ is considered entangled when $\pi_{SB}$ cannot be decomposed into a classical mixture of tensor product states [the right-hand side of Eq.~\eqref{statebased}].
\end{definition}
This corresponds to entanglement with respect to the $S_{0\pi}$ universality class from Ref.~\cite{banuls2007}. The physical meaning of this definition becomes clear when noting that the state $\pi_{SB}$ reproduces all the correlations of local observables $O_S$ and $O_B$ that act on the system and the bath,
\begin{align}
	\forall O_S O_B: \quad \text{Tr} \left[O_S O_B \rho_{SB} \right] = \text{Tr} \left[O_S O_B \pi_{AB} \right],
\end{align}
as the observables obey the parity superselection rule. Therefore, the state $\rho_{SB}$ is deemed separable when it cannot be distinguished from a classical mixture of tensor product states via correlations of local measurements (e.g., through a violation of Bell's inequality).

\subsection{Entanglement witnessing} \label{subsec:entwitn}
Let us now present how the presence of entanglement with respect to Definition~\ref{obsbased} can be detected. As discussed in Ref.~\cite{ding2022}, this is generally a nontrivial task. However, as shown by Ba\~{n}uls \textit{et al.}~\cite{banuls2007}, the entanglement witnessing becomes simple for states which can be represented as a tensor product of two identical copies of the system-bath density matrix: $\hat{\rho}_{SB}=\rho_{SB} \otimes \rho_{SB}$. Then the total state of both copies $\hat{\rho}_{SB}$ is entangled if and only if the partially transposed density matrix of a single copy $\rho_{SB}^{T_B}$ is negatively defined (i.e., it has some negative eigenvalues). Here $T_B$ denotes the partial transposition of the bath state; for its definition, see Refs.~\cite{peres1996, horodecki1996} and the Appendix~\ref{sec:transp}. Entanglement can then be witnessed by the positivity of the entanglement negativity~\cite{eisert2001, vidal2002}
\begin{align} \label{negativity}
		\mathcal{N}=\sum_{\lambda_i <0} |\lambda_i|=\sum_i \frac{|\lambda_i|-\lambda_i}{2},
	\end{align}
where $\lambda_i$ are the eigenvalues of $\rho_{SB}^{T_B}$. As one may observe, the entanglement negativity by construction exhibits a nonanalytic behavior, being equal to 0 for separable states and taking positive values for the entangled states. We further note that in fermionic systems positivity of the entanglement negativity is a \textit{necessary} and sufficient condition of entanglement~\cite{banuls2007}, while in a generic case it is only a sufficient condition~\cite{peres1996, horodecki1996}.

The scenario described above corresponds to a situation in which we have two identical copies of a fermionic system. As a physically relevant example, one may consider a spinful system being a thermal state of a spin-degenerate quadratic Hamiltonian $\hat{H}=\sum_{ij} \sum_{\sigma \in \{\uparrow,\downarrow\}} \mathcal{A}_{ij} c_{i\sigma}^\dagger c_{j\sigma}$, or evolving under such a Hamiltonian. Then the total density matrix can be represented as a tensor product of two identical density matrices corresponding to subspaces of spin $\uparrow$ and $\downarrow$ levels: $\hat{\rho}_{SB}=\rho_{SB}^\uparrow \otimes \rho_{SB}^\downarrow$ with $\rho_{SB}^\uparrow=\rho_{SB}^\downarrow$. In the following discussion we will always assume the presence of two identical copies of $\rho_{SB}$, without assuming any specific physical realization.

We further note that we apply a standard definition of negativity used in the field of quantum information~\cite{eisert2001, vidal2002} rather than the ``fermionic negativity'' defined in Ref.~\cite{shapourian2017} and later used in several studies~\cite{gruber2020, murciano2021, turkeshi2022, eisler2022, fraenkel2022, fraenkel2023}. This is because, as shown in Ref.~\cite{shapourian2019}, the latter quantity witnesses entanglement with respect to the $S_{2\pi}$ equivalence class from Ref.~\cite{banuls2007}, which is a much weaker notion than Definition~\ref{obsbased}. In particular, for $1+N$-mode fermionic Gaussian states entanglement with respect to the $S_{2\pi}$ equivalence class is equivalent to the presence of correlations~\cite{banuls2007, spee2018}, and thus the ``fermionic negativity'' is positive for every correlated state. In contrast, entanglement with respect to Definition~\ref{obsbased} appears only when the correlations reach a certain finite threshold~\cite{banuls2007}.

\section{Model and methods} \label{sec:model}
\subsection{Noninteracting resonant level model} \label{sec:nonint}
Let us now present the details of the considered model and the methods we use to characterize the system-bath entanglement. The paper focuses on the noninteracting resonant level model consisting of a single fermionic energy level tunnel-coupled to a fermionic bath with the inverse temperature $\beta=1/(k_B T)$ and the chemical potential $\mu$. Generalization to the case of multiple baths will be considered in Sec.~\ref{sec:volt}. It is described by the Hamiltonian
		\begin{align} \label{hamnrl}
	\hat{H}=\epsilon_{0} c^\dagger_{0} c_{0} +\sum_{k=1}^K \epsilon_{k} c_{k}^\dagger c_{k} + \sum_{k=1}^K \left( t_k c^\dagger_{0} c_{k} + \text{h.c.} \right),
\end{align}
where the index $k=0$ corresponds to the system, while $k \in \{1,\ldots,K\}$ to the energy levels of the bath. Here $\epsilon_k$ is the level energy, $c^\dagger_{k}$ and $c_{k}$ are the creation and annihilation operators, $t_k$ is the tunnel coupling between the levels $0$ and $k$, and $K$ is the number of energy levels in the bath. The Hamiltonian is taken to be spinless. A possible role played by spin was discussed in Sec.~\ref{subsec:entwitn} in the context of entanglement witnessing.

To fix the parameters of the Hamiltonian, we now use the following convention. In theory of open quantum systems, the bath is characterized by means of its spectral density $\Gamma(\omega)=\sum_k 2 \pi \delta (\epsilon_k-\omega) |t_k|^2$~\cite{schaller2014}. In the continuous limit of infinitesimal level spacing this can be rewritten as $\Gamma(\epsilon_k)=2 \pi |t_k|^2 \xi(\epsilon_k)$, where $\xi(\omega)$ is the density of states in the bath. We now focus on a boxcar-shaped spectral density defined as
\begin{align}
	\Gamma(\omega)=	\begin{cases}
		\Gamma \quad \text{for} \quad \omega \in [-W/2,W/2], \\
		0  \quad \text{otherwise},
	\end{cases}
\end{align}
where $W$ is the bandwidth. Accordingly, we later consider a discretized version of this model in which we uniformly distribute the energy levels of the bath $\epsilon_k$ throughout the interval $[-W/2,W/2]$, and parameterize the tunnel couplings as $\Gamma={2 \pi t_k^2 (K-1)/W}$, where we take $t_k$ to be positive real numbers.

\subsection{Correlation matrix approach} \label{subsec:cormat}
To evaluate the entanglement between the system and the macroscopic bath, it is necessary to know the total system-bath state $\rho_{SB}$. This may seem infeasible, since the dimension of the Hilbert space increases exponentially with the number of levels in the bath $K$. However, this problem can be circumvented for noninteracting systems described by the quadratic Hamiltonian, such as Eq.~\eqref{hamnrl}. Indeed, then (for the grand canonical thermal state, or evolution starting from such a state) the total state is Gaussian, which means that it is fully described by the $(1+K) \times (1+K)$ correlation matrix $\mathcal{C}_{kl} ={\text{Tr} (\rho_{SB} c_k^\dagger c_l)}$~\cite{peschel2003}. Its evolution follows the equation~\cite{eisler2012}
\begin{align} \label{cormatdyn}
	\mathcal{C}(t)=e^{i\mathcal{H}t} \mathcal{C}(0) e^{-i\mathcal{H}t},
\end{align} 
where $\mathcal{H}$ is the single-particle Hamiltonian defined as
\begin{align} \label{fermhamsp}
	\begin{cases}
		\mathcal{H}_{kk}= \epsilon_k & \text{for} \quad k=0,\ldots,K, \\
		\mathcal{H}_{0k}=\mathcal{H}_{k0}=t_k & \text{for} \quad k=1,\ldots,K, \\
		\mathcal{H}_{kl}=0 &  \text{otherwise}.
	\end{cases}
\end{align}
Here and from hereon we take $\hbar=1$. The initial correlation matrix $\mathcal{C}(0)$ reads
\begin{align}
	\mathcal{C}(0) = \left[p_0,f(\epsilon_1),\ldots,f(\epsilon_K) \right],
\end{align}
where $p_0$ is the initial occupancy of the system and ${f(\epsilon)=1/\{1+\exp[\beta(\epsilon-\mu)]\}}$ is the Fermi distribution.

\subsection{Calculation of the entanglement negativity} \label{subsec:nunnegcalc}
To calculate the entanglement negativity one still needs the density matrix rather than the correlation matrix. This is because (in contrast to bosonic systems), even when $\rho_{SB}$ is a Gaussian state, the partially transposed state $\rho_{SB}^{T_B}$ is not a Gaussian operator~\cite{eisler2015}. The density matrix can be obtained from the correlation matrix as~\cite{peschel2003}
\begin{align} \label{denmatfromcor}
	\rho_{SB}=\frac{\exp(-\sum_{kl} \mathcal{B}_{kl} c_k^\dagger c_l)}{\text{Tr} \exp(-\sum_{kl} \mathcal{B}_{kl} c_k^\dagger c_l)},
\end{align}
where $\mathcal{B} = \ln [(\mathds{1}-\mathcal{C})\mathcal{C}^{-1}]$. The creation and annihilation operators can be expressed in a matrix form, e.g., by means of the Jordan-Wigner transform. Unfortunately, as mentioned above, the calculation of the full density matrix is unfeasible for large baths, as its size grows exponentially with $K$. To deal with this obstacle, we use the following approach. First, we put the correlation matrix $\mathcal{C}$ into the tridiagonal form by means of the Householder transformation~\cite{householder1958} (see Ref.~\cite{ozaki} and the Appendix~\ref{sec:hous} for details of the algorithm used). As such a transformation is realized by a unitary operation acting on the bath only, it does not change the entanglement negativity. Then, we calculate the ``partial'' entanglement negativity $\mathcal{N}_M$ between the system and the part of the bath consisting of the first few fermionic modes $i=1,\ldots,M$. Due to the monotonicity property~\cite{eisert2001, vidal2002}, this quantity provides a lower bound for the total negativity: $\mathcal{N} \geq \mathcal{N}_M$. Later, we mostly apply the cutoff $M=6$, which we found to be sufficient to provide a good estimate of the total negativity $\mathcal{N}$ in most of the considered parameter regimes; this will be illustrated on a specific example in Fig.~\ref{fig:cutoff}.

\section{Equilibrium entanglement} \label{sec:equi}
Let us now present the results. In this section we investigate entanglement in the global equilibrium state of the system and the bath, focusing on two distinct thermodynamic scenarios where the joint system-bath state is described by either the grand canonical ensemble with a fluctuating particle number (Sec.~\ref{sec:eqgrand}), or the canonical ensemble with a fixed particle number (Sec.~\ref{sec:eqcan}). As will be demonstrated, although both ensembles provide the same reduces state of the system (in the thermodynamic limit), they lead to both quantitatively and qualitatively different behavior of the system-bath entanglement.

\subsection{Grand canonical ensemble} \label{sec:eqgrand}
In the first step, we analyze entanglement between the system and the bath for the grand canonical state of the total Hamiltonian
\begin{align}
	\rho_{SB}=\frac{e^{-\beta(\hat{H}-\mu \hat{N})}}{\text{Tr} [e^{-\beta(\hat{H}-\mu \hat{N})}]},
\end{align}
where $\hat{N}$ is the particle number operator. As the grand canonical Gibbs state of a quadratic Hamiltonian is Gaussian, we apply the correlation matrix approach. We determine the correlation matrix in the following way. First, the single-particle Hamiltonian $\mathcal{H}$ defined by Eq.~\eqref{fermhamsp} is diagonalized as
\begin{align}
	\mathcal{H}=\mathcal{P} \mathcal{H}^D \mathcal{P}^\dagger,
\end{align}
where $\mathcal{H}^D$ is a diagonal matrix. Then the correlation matrix of the global Gibbs state $\mathcal{C}^{\text{eq}}$ can be calculated as
\begin{align}
	\mathcal{C}^{\text{eq}} = \mathcal{P} \mathcal{C}^{\text{eq},D} \mathcal{P}^\dagger,
\end{align}
where $\mathcal{C}^{\text{eq},D}$ is the equilibrium correlation matrix expressed in the basis diagonalizing $\mathcal{H}$. Explicitly, it is expressed as $\mathcal{C}^{\text{eq},D}=\text{diag}[f(\mathcal{H}^D_{00}),\ldots,f(\mathcal{H}^D_{KK})]$, where, to recall, $f(\epsilon)$ is the Fermi distribution. The entanglement negativity can then be calculated using the method presented in Sec.~\ref{subsec:nunnegcalc}.

We now analyze the entanglement behavior for different system parameters. Let us clarify the unit convention that we use. In most cases, the entanglement will be plotted as a function of the non-dimensional ratio $\Gamma/(k_B T)$. This may be interpreted either as a function of the coupling strength $\Gamma$ for a constant temperature, or as a function of the inverse temperature $1/T$ for a constant $\Gamma$. Since both parameters are tunable in experiments~\cite{cronenwett1998, svilans2016}, both interpretations are physically meaningful. The rest of the parameters will be expressed in units of $\Gamma$ or $k_B T$, such that the results do not change when both $\Gamma$ and $k_B T$ are multiplied by the same factor. 

\begin{figure}
	\centering
	\includegraphics[width=0.9\linewidth]{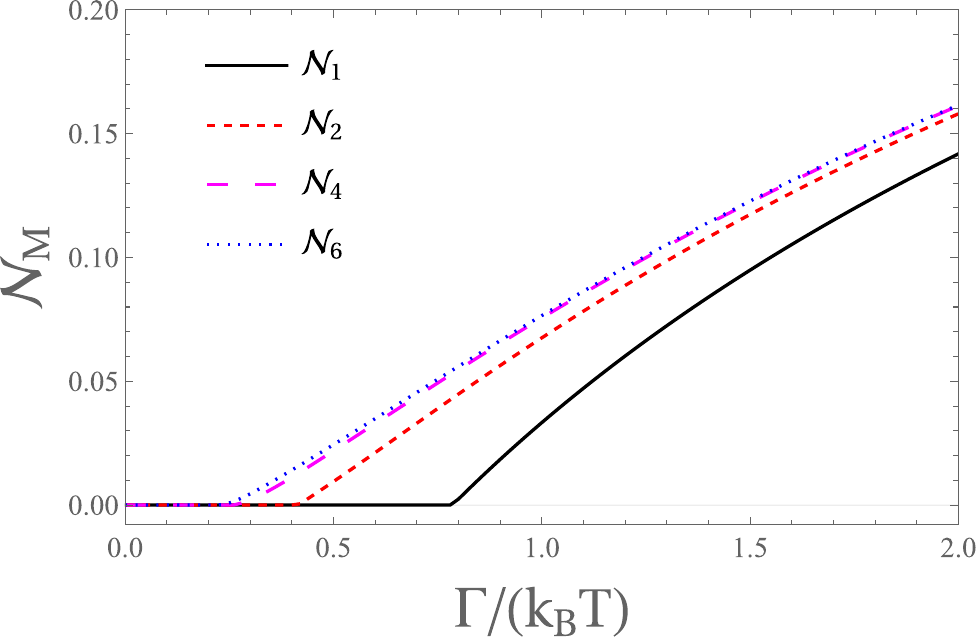}
	\caption{Entanglement negativity $\mathcal{N}_M$ as a function of the coupling strength $\Gamma$ for different values of the cutoff $M$. Results for $\epsilon_0=\mu=0$, $W=50 \Gamma$, and $K=400$.}
	\label{fig:cutoff}
\end{figure}

In the first step, we investigate the behavior of partial negativities $\mathcal{N}_M$ for different values of the cutoff $M$ to determine whether they provide a good estimate of the total negativity $\mathcal{N}$. The results are presented in Fig.~\ref{fig:cutoff}. As may be noted, for all partial negativities, entanglement is absent for weak coupling strengths, but appears for a finite value of $\Gamma$ of the order of magnitude of the thermal energy $k_B T$. This is because for mixed states entanglement appears only when system-bath correlations (which gradually build up when the coupling strength increases) reach a certain finite threshold~\cite{zyczkowski2001, yu2009}; this contrasts with the behavior of pure states, where every correlated state is entangled. The threshold value of $\Gamma/(k_B T)$, at which entanglement appears, decreases with increasing cutoff $M$. However, the partial negativities $\mathcal{N}_4$ and $\mathcal{N}_6$ are already very close to each other. This suggests that $\mathcal{N}_6$ provides a good estimate of the total negativity $\mathcal{N}$. From hereon, we mostly apply this value of the cutoff.

\begin{figure}
	\centering
	\includegraphics[width=0.9\linewidth]{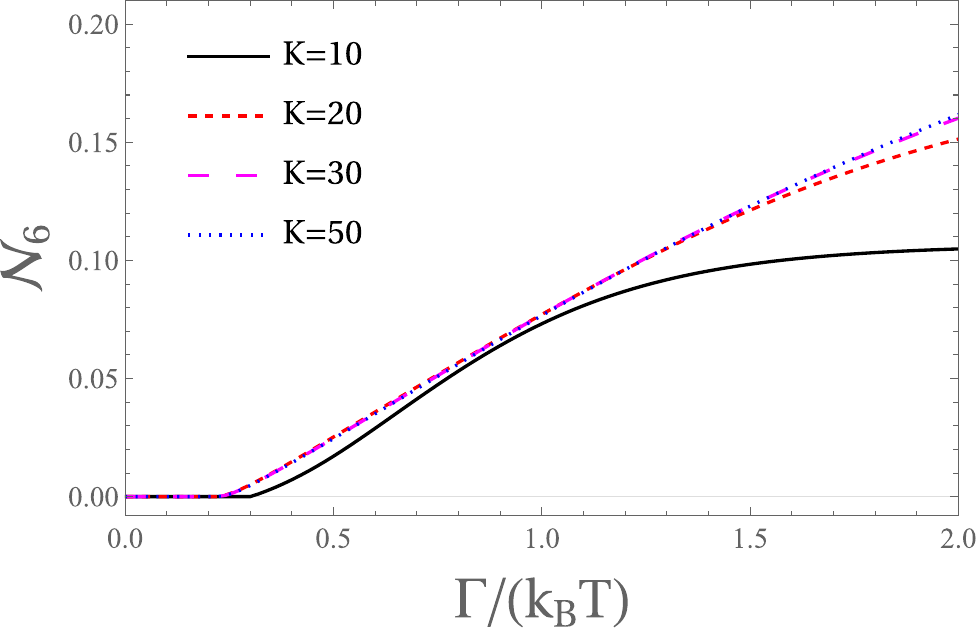}
	\caption{Entanglement negativity $\mathcal{N}_6$ for different numbers of bath levels $K$. Other parameters as in Fig.~\ref{fig:cutoff}.}
	\label{fig:eqkdep}
\end{figure}

In Fig.~\ref{fig:eqkdep} we analyze the dependence of entanglement on the number of bath levels $K$ (and thus on the density of states in the bath). Apart from the fundamental importance for real finite systems, this is an important technical parameter. In numerical simulations we can only deal with finite baths; therefore, it is necessary to establish whether they can adequately simulate the thermodynamic limit. As can be observed, for small $K=\{10,20,30\}$ entanglement increases with the bath size. However, for larger baths the entanglement negativity becomes nearly size-independent; indeed, the results for $K=30$ and $K=50$ are already very close to each other. On this basis, we can conclude that a sufficiently large finite bath can adequately simulate the thermodynamic limit. Specifically, to reproduce the thermodynamic limit, the interlevel spacing in the bath $\Delta \epsilon= W/K$ must be approximately smaller than $\Gamma$.

\begin{figure}
	\centering
	\includegraphics[width=0.9\linewidth]{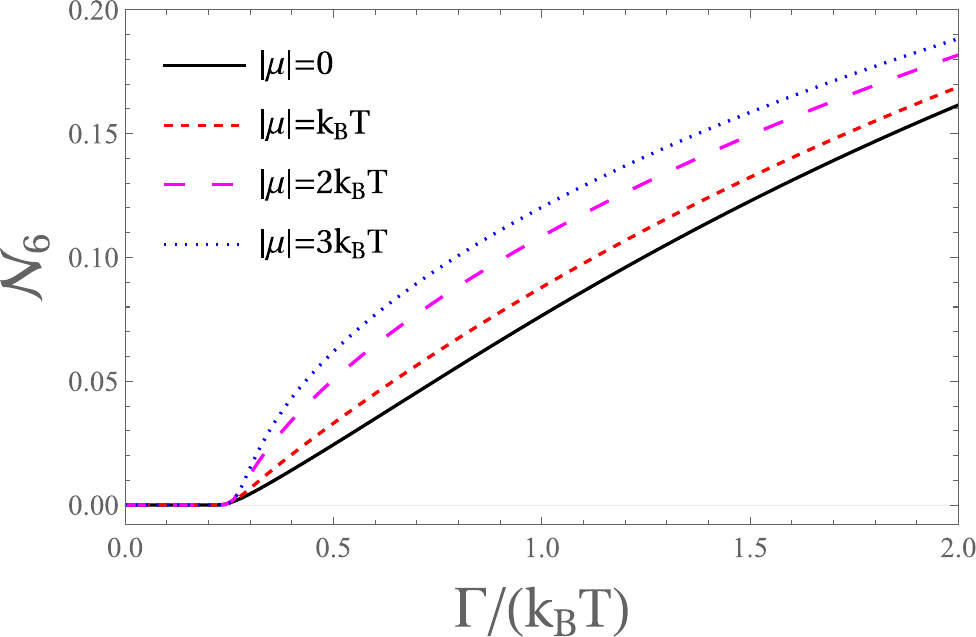}
	\caption{Entanglement negativity $\mathcal{N}_6$ for different values of the chemical potential $\mu$. Other parameters as in Fig.~\ref{fig:cutoff}.}
	\label{fig:eqgdep}
\end{figure}

In Fig.~\ref{fig:eqgdep} we further investigate the dependence of entanglement on the chemical potential $\mu$, and thus on the degree of breaking of the particle-hole symmetry. As shown, for all values of $\mu$, entanglement appears at the same threshold value of $\Gamma/(k_B T)$. However, its magnitude increases with the absolute value of the chemical potential. This may be related to an increase in the purity of the bath levels that are resonant with the system.

\begin{figure}
	\centering
	\includegraphics[width=0.9\linewidth]{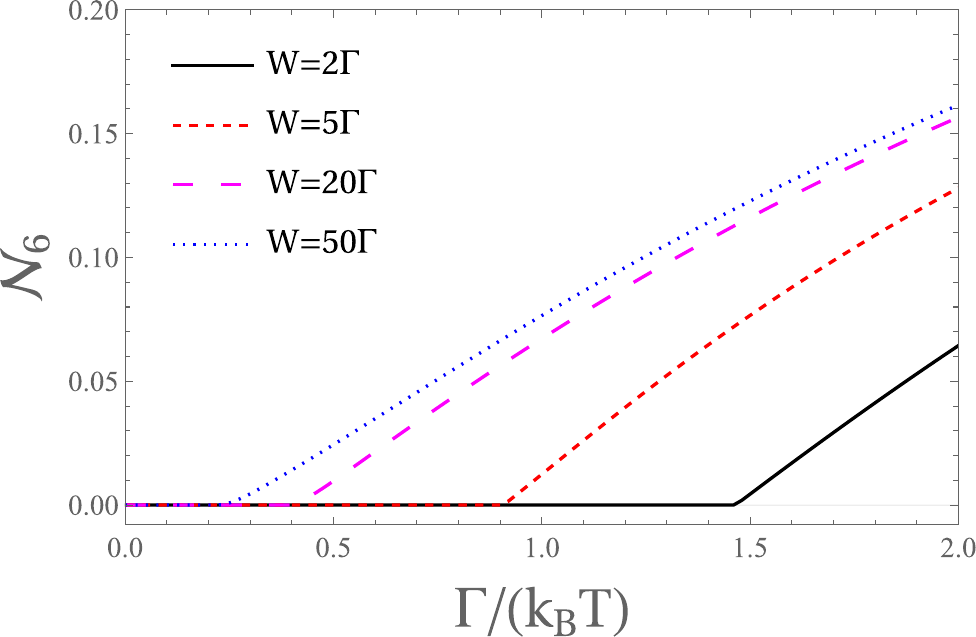}
	\caption{Entanglement negativity $\mathcal{N}_6$ for different values of the bandwidth $W$, with $K=8W/\Gamma$. Other parameters as in Fig.~\ref{fig:cutoff}.}
	\label{fig:entband}
\end{figure}

\begin{figure}
	\centering
	\includegraphics[width=0.9\linewidth]{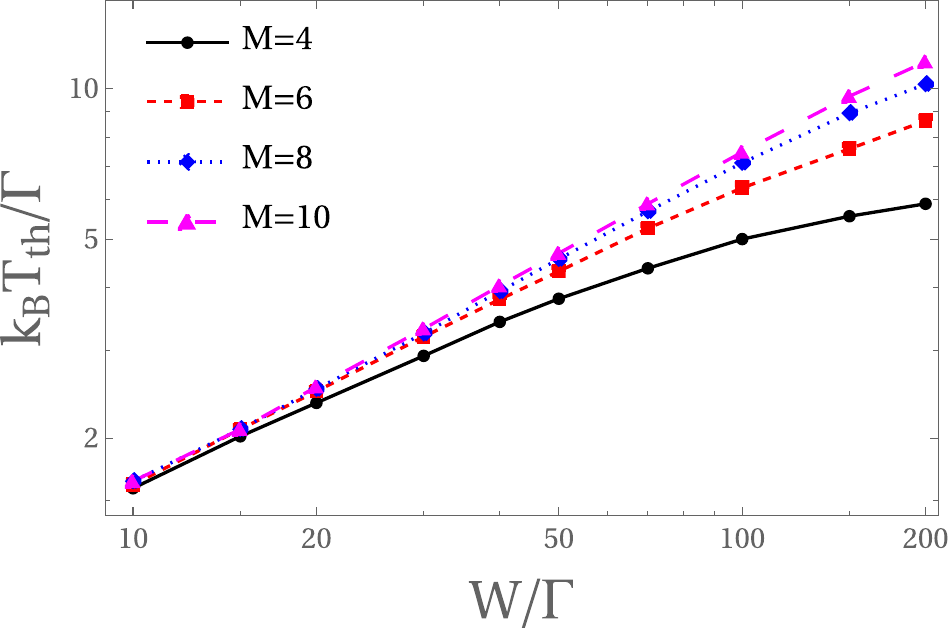}
	\caption{Threshold temperature $T_\text{th}$, below which the entanglement is present, as a function of the bandwidth $W$, evaluated with different cutoffs $M$. Parameters as in Fig.~\ref{fig:cutoff}, unless denoted otherwise.}
	\label{fig:tth}
\end{figure}

Finally, we analyze how entanglement depends on the bath bandwidth $W$. To keep the density of states in the bath constant, we take the number of bath levels $K$ to be proportional to the bandwidth. The results are presented in Fig.~\ref{fig:entband}. Since the bandwidth is usually not tunable, one may interpret the figure as plotted as a function of the inverse temperature $1/T$ for constant $\Gamma$ and $W$. As shown, for smaller bandwidths the entanglement negativity is also smaller and appears at higher values of $\Gamma/(k_B T)$. Interestingly, one can observe a pronounced difference between the entanglement behavior for bandwidths larger than the coupling strength by an order of magnitude ($W=20 \Gamma$ and $W=50\Gamma$). This result suggests that entanglement may be strongly affected by details of the spectral density of the bath $\Gamma(\omega)$, even for energies $\omega$ far from resonance with the system energy $\epsilon_0$. To illustrate this further, in Fig.~\ref{fig:tth} we plot (on a log-log scale) the threshold temperature $T_\text{th}$, below which the entanglement is present, as a function of $W/\Gamma$. It is evaluated for different cutoffs $M$. As one can first note, for large bandwidths one needs to use large cutoffs to make the calculations reliable. As a consequence, our simulations are limited to $W \leq 200\Gamma$. Second, the threshold temperature increases monotonically as the bandwidth increases. In particular, in the bandwidth range considered, the threshold temperature evaluated for the cutoff $M=10$ obeys approximately a power law $T_\text{th} \propto \Gamma (W/\Gamma)^{2/3}$. This might suggest that in the infinite bandwidth limit the entanglement appears for any finite temperature and coupling strength $\Gamma$. However, as our calculations are limited to finite bandwidths, and naive extrapolations of a finite-size scaling are sometimes misleading, it is not possible to state it with certainty.

This result might be surprising, as intuitively the strongly off-resonant levels of the bath should be very weakly correlated with the system. A possible explanation of this phenomenon may be provided by considering the thermal entanglement in a toy model of two coupled fermionic levels. It is described by the Hamiltonian
\begin{align} \label{hamdet}
	\hat{H}=\epsilon_1 c_1^\dagger c_1+\epsilon_2 c_2^\dagger c_2 + \mathcal{T} (c_1^\dagger c_2 + c_2^\dagger c_1),
\end{align}
where $\epsilon_1$ and $\epsilon_2$ are the level energies, and $\mathcal{T}$ is the tunnel coupling. Using Jordan-Wigner transform, it can be expressed in a matrix form as
\begin{align} \label{hamdetmat}
\hat{H}=\begin{pmatrix}
	\epsilon_1+\epsilon_2 & 0 & 0 & 0 \\
	0 & \epsilon_1 & \mathcal{T} & 0 \\
	0 & \mathcal{T} & \epsilon_2 & 0 \\
	0 & 0 & 0 & 0
\end{pmatrix}.
\end{align}
\begin{figure}
	\centering
	\includegraphics[width=0.9\linewidth]{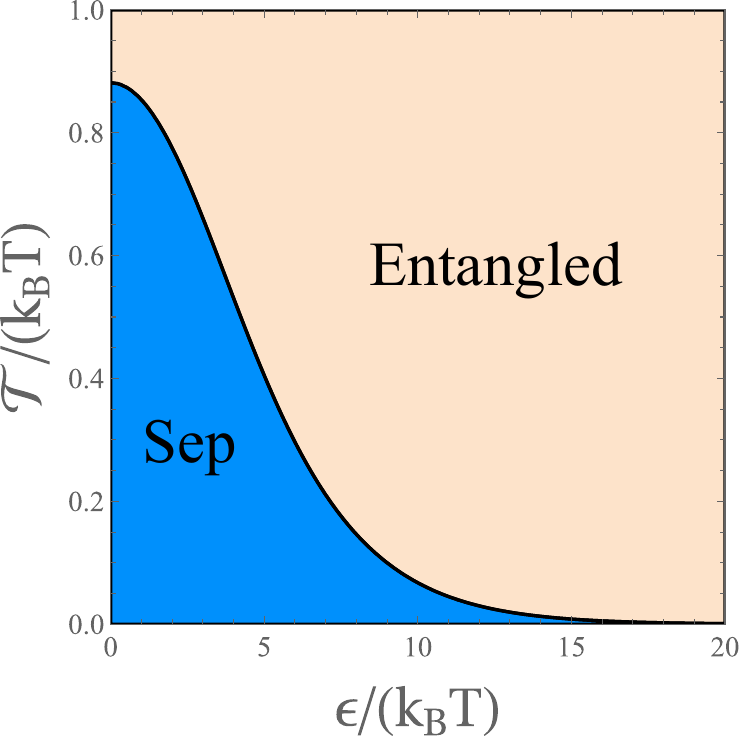}
	\caption{Entanglement phase diagram for two fermionic levels as a function of the level detuning $\epsilon$ and the tunnel coupling $\mathcal{T}$. ``Sep'' denotes the separable phase.}
	\label{fig:phasediagdet}
\end{figure}
For such a model, a density matrix of the thermal state $\rho=Z^{-1} \exp[-\beta(\hat{H}-\mu \hat{N})]$, and thus the entanglement negativity, can be evaluated explicitly. Let us now take $\epsilon_1=\mu=0$ and $\epsilon_2=\epsilon$; the parameter $\epsilon$ describes then the detuning of the energy levels. In Fig.~\ref{fig:phasediagdet} we present the phase diagram of the thermal entanglement as a function of the level detuning and the tunnel coupling. As may be noted, for a larger detuning $\epsilon$, the entanglement appears at lower threshold values of the tunnel coupling $\mathcal{T}$. In the limit of $\epsilon \rightarrow \infty$, the entanglement is present for any finite $\mathcal{T}$. Thus, off-resonant fermionic levels are more liable to be entangled than the resonant ones. A similar occurrence has been previously observed for qubits in an inhomogeneous magnetic field~\cite{asoudeh2005, zhang2005}.

\begin{figure}
	\centering
	\includegraphics[width=0.9\linewidth]{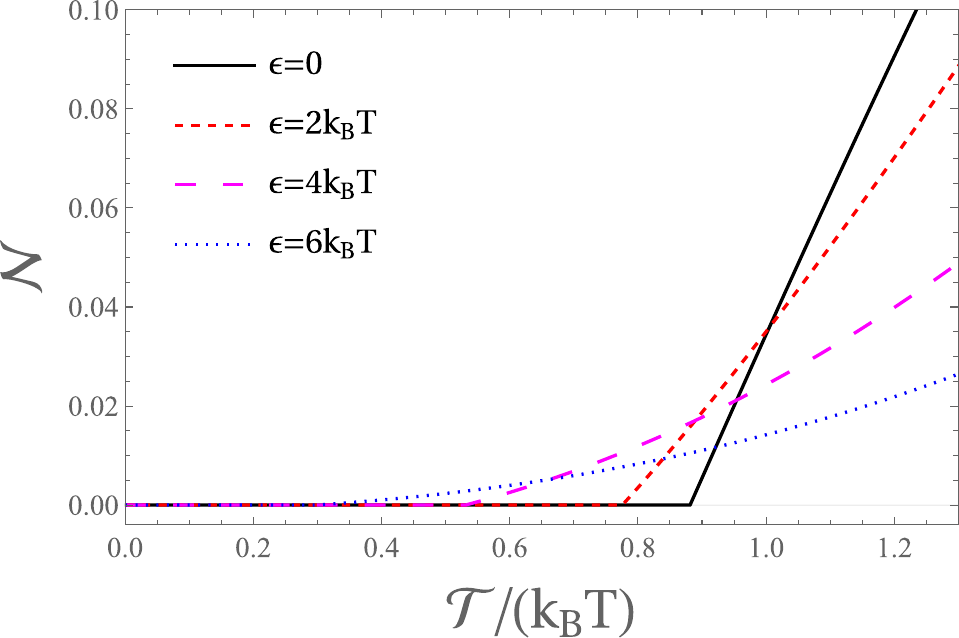}
	\caption{Entanglement negativity in a two-level system as a function of the tunnel coupling $\mathcal{T}$ for different level detunings $\epsilon$.}
	\label{fig:ent2lev}
\end{figure}
This results is still not intuitive, as the off-resonant levels should be more weakly correlated than the resonant ones. Indeed, as shown in Fig.~\ref{fig:ent2lev}, while the level detuning decreases the threshold tunnel coupling, it also quantitatively suppresses the entanglement for large $\mathcal{T}$. To provide a qualitative explanation of this phenomenon, let us consider a perturbative form of the density matrix $\rho$ for the case of a large detuning $\epsilon$ and a small tunnel coupling $\mathcal{T}$. To this end, we treat a diagonal part of the Hamiltonian~\eqref{hamdetmat} as an unperturbed Hamiltonian, and the off-diagonal part as a small perturbation. The density matrix can then be approximated as
\begin{align}
	\rho \approx Z^{-1} \sum_{i=1}^4 |\psi_i^{(1)} \rangle \langle \psi_i^{(1)} | e^{-\beta E_i^{(1)}},
\end{align} 
where $|\psi_i^{(1)} \rangle$ are eigenstates of Eq.~\eqref{hamdet} with energies $E_i^{(1)}$ obtained within first-order perturbation theory. In the limit of $\epsilon \gg k_B T \gg \mathcal{T}$ this yields
\begin{align}
	\rho \approx \frac{1}{2} \begin{pmatrix}
		e^{-\beta \epsilon} & 0 & 0 & 0 \\
		0 & 1 & -\mathcal{T}/\epsilon & 0 \\
		0 & -\mathcal{T}/\epsilon & \mathcal{T}^2/\epsilon^2+e^{-\beta \epsilon} & 0 \\
		0 & 0 & 0 & 1
	\end{pmatrix}.
\end{align}
The partially-transposed density matrix takes then the form
\begin{align}
	\rho^{T_B} \approx \frac{1}{2} \begin{pmatrix}
		e^{-\beta \epsilon} & 0 & 0 & -\mathcal{T}/\epsilon \\
		0 & 1 & 0 & 0 \\
		0 & 0 & \mathcal{T}^2/\epsilon^2+e^{-\beta \epsilon} & 0 \\
		-\mathcal{T}/\epsilon & 0 & 0 & 1
	\end{pmatrix}.
\end{align}
The matrix $\rho^{T_B}$ is non-positively defined, and thus the system is entangled, for $\mathcal{T} \gtrapprox \epsilon e^{-\beta \epsilon/2}$. One can numerically check that this approximation works well for $\epsilon \gtrapprox 3 k_B T$.

This analytic result leads us to a qualitative explanation of the reduced tunnel coupling threshold for a large detuning. While increasing the detuning suppresses the correlation between levels 1 and 2, it also increases the purity of the state of level 2 by reducing its occupancy. Thus, even though correlations between levels are weaker, they are more likely to be genuinely quantum because of the increased purity. Furthermore, while the off-diagonal elements of the density matrix (related to the interlevel coherence) decay only algebraically with the detuning as $\mathcal{T}/\epsilon$, the occupancy of the level $2$ decreases exponentially as $e^{-\beta \epsilon}$. Thus, the effect of the increased purity is stronger than that of the reduced correlation, which promotes the presence of entanglement. 

Based on this, we may try to provide an explanation for the observed bandwidth dependence: For a large bandwidth, the system is coupled to a large number of strongly detuned levels in the bath. Although the system is only weakly correlated with them, they might still significantly contribute to entanglement, since their states are highly pure (with occupancy close to either 0 or 1), as illustrated by the two-level model.

We finally note that our result may be important for numerical simulations of the system-bath entanglement in strongly correlated impurities, which recently gained notable attention~\cite{lee2015, shim2018, kim2021, bayat2010, wagner2018, shim2023, alkurtass2016, bayat2017, bayat2012, bayat2014, yoo2018, mihailescu2022}. In simulations, the bandwidth is often treated as a technical parameter that should be kept larger than other energy scales of the system to avoid its influence on the system behavior. While this approach is often valid when considering the system observables, our results suggest that one must be more careful in the case of information-theoretic correlations, such as the system-bath entanglement. Then, the effect of the bandwidth can still be important, even when it significantly exceeds other energy scales.

\subsection{Canonical ensemble} \label{sec:eqcan}
In the previous subsection, the equilibrium state of the system and the bath was described within the grand canonical ensemble, with fluctuating energy and particle number. However, within the framework of statistical physics, alternative choices can be considered, such as the canonical ensemble with a fixed particle number. Indeed, the latter choice may appear to be more physically justified for certain physical setups, such as impurities interacting with a trapped cloud of ultracold atoms~\cite{bauer2013}, where the number of particles in the experimental setup is fixed. As follows from the principle of ensemble equivalence, in the thermodynamic limit both ensembles predict the same reduced state of the system. However, as shown in our previous paper~\cite{ptaszynski2023ens}, the transient properties of microscopic system-bath correlations may depend on the choice of the ensemble. Here we show that this is also true for the equilibrium entanglement.

Since the canonical state with a fixed particle number is not a Gaussian state, one needs to operate on the level of full density matrices rather than use the correlation matrix approach. The density matrix for the canonical state with the particle number $N$ is calculated as
\begin{align}
\rho_{SB} = Z^{-1} \sum_{i} \delta_{N_i,N} e^{-\beta E_i} |\psi_i \rangle \langle \psi_i|,
\end{align}
where $|\psi_i \rangle$ is the eigenstate of the total Hamiltonian $\hat{H}$ with energy $E_i$ and particle number $N_i$, while $Z=\sum_{i} \delta_{N_i,N} e^{-\beta E_i}$ is the partition function. Due to the need to calculate the full density matrix, we choose a small number of bath levels $K=7$ or $K=9$. We also take a relatively small bandwidth $W=5 \Gamma$. This is a sort of compromise: While for small bandwidths the model does not reproduce the properties of typical wide-band baths considered in the literature, for large bandwidths the energy levels are no longer sufficiently dense to simulate the continuous spectral density.

\begin{figure}
	\centering
	\includegraphics[width=0.9\linewidth]{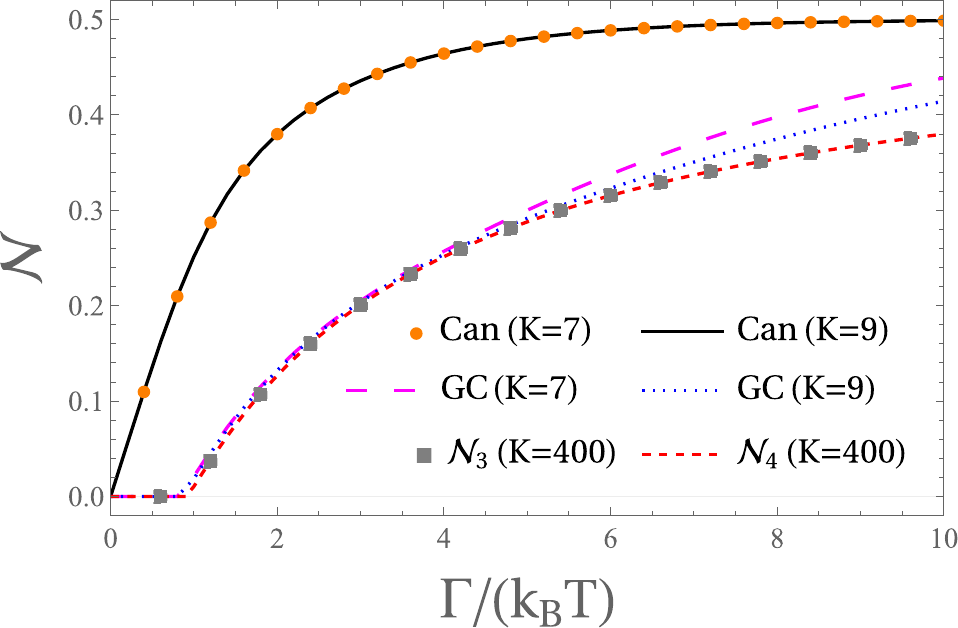}
	\caption{Entanglement negativity $\mathcal{N}$ as a function of the coupling strength $\Gamma$ at the particle-hole symmetric point $\epsilon_0=\mu=0$ for the canonical (Can) and the grand canonical (GC) ensembles for the number of bath levels $K=7$ and $K=9$, and the bandwidth $W=5 \Gamma$. Results compared with the partial negativities $\mathcal{N}_3$ and $\mathcal{N}_4$ calculated for the grand canonical ensemble with $K=400$.}
	\label{fig:eqens}
\end{figure}

The entanglement negativity calculated for different ensembles is presented in Fig.~\ref{fig:eqens}. First, as in the previous section, for the grand canonical ensemble entanglement appears for a finite value of the coupling strength $\Gamma$ of the order of $k_B T$. As shown by comparison with the partial negativities $\mathcal{N}_3$ and $\mathcal{N}_4$ calculated using the correlation matrix approach for $K=400$, the threshold value is not affected by the small size of the bath. Indeed, for a sufficiently small $\Gamma \lessapprox 4 k_B T$ the results coincide, suggesting that the considered small baths with $K=7$ or $K=9$ levels already reproduce the properties of entanglement in the thermodynamic limit. For larger $\Gamma \gtrapprox 4 k_B T$ the entanglement negativity depends on $K$ more strongly, decreasing with the number of levels, and approaching the value of partial negativities $\mathcal{N}_3$ and $\mathcal{N}_4$ calculated for a large bath $K=400$; the latter quantities coincide, which suggests that they appropriately approximate the total entanglement negativity for large baths.

In contrast, the canonical ensemble entanglement is present for arbitrarily weak finite coupling strengths $\Gamma$ (given the finite temperature $T$). This can be explained as follows. Let us first define the many-particle Fock states $|\phi_\mathbf{y} \rangle=(c_K^\dagger)^{y_K} \ldots (c_0^\dagger)^{y_0} |\varnothing \rangle$, where $\mathbf{y}=(y_0,\ldots,y_K)$ is the vector of level occupancies and $|\varnothing \rangle$ is the vacuum state. As one can note, the Fock states are characterized with a definite number of particles in each level. Then, according to the theory presented in Ref.~\cite{ma2022}, for a fixed particle number entanglement is present whenever there exist nonzero off-diagonal elements of the density matrix expressed in the Fock basis $(\rho_{SB})_{\mathbf{yz}}=\langle \phi_\mathbf{y}|\rho_{SB}|\phi_\mathbf{z} \rangle$, with $|\phi_\mathbf{y} \rangle$ and $|\phi_\mathbf{z} \rangle$ corresponding to different occupancies of the system $S$ (i.e., $y_0 \neq z_0$). Such off-diagonal elements are obviously present in the thermal state of the Hamiltonian~\eqref{hamnrl}, which is not diagonal in the Fock basis; this is due to the presence of the tunneling term $\sum_{k=1}^K ( t_k c^\dagger_{0 \sigma} c_{k \sigma} + \text{h.c.})$, which coherently exchanges particles between the system and the bath. Furthermore, the entanglement negativity for the canonical state significantly exceeds the one calculated for the grand canonical ensemble, although both converge to the asymptotic value $1/2$ in the limit $\Gamma/(k_B T) \rightarrow \infty$. This demonstrates that the system-bath entanglement depends on the statistical ensemble describing the global equilibrium state.

\section{Transient dynamics}
In this section we analyze the entanglement generated during the transient relaxation of a fermionic impurity initialized out-of-equilibrium with respect to the bath. First, in Sec.~\ref{sec:antheor} we present an analytic theory applicable to weakly coupled impurities. In Sec.~\ref{sec:numres} we present the numerical results establishing a range of applicability of this theory, as well as providing insight into the transient entanglement behavior in the strong-coupling regime.

\subsection{Analytic theory} \label{sec:antheor}
\subsubsection{Derivation} \label{subsec:andert}
Here we present an analytic theory enabling to calculate the entanglement negativity in the regime of weak system-bath coupling $\Gamma \ll k_B T$. The method used is based on reconversion of multimode into two-mode correlations via a suitable unitary operation acting of the bath. This approach was first proposed by Botero and Reznik for bosonic Gaussian states~\cite{botero2003}, and later thoroughly investigated in Refs.~\cite{adesso2004, serafini2005}.

Our theory is based on the following reasoning. The energy level of the system is effectively (resonantly) coupled only to those energy levels in the bath, whose energies $\epsilon_k$ are close to the energy of the system $\epsilon_0$, that is, the interlevel separation $|\epsilon_k-\epsilon_0|$ is of the order of the level broadening $\Gamma$. When the coupling strength to the bath is weak compared to temperature ($\Gamma \ll k_B T$), the occupancies of these levels may be approximated by the Fermi distribution at $\epsilon_0$: $f(\epsilon_k) \approx f(\epsilon_0)$. One may thus consider an initial state of the bath where all levels of the bath have an initial occupancy $f=f(\epsilon_0)$. One must be aware that -- as illustrated by the bandwidth dependence of entanglement in the equilibrium case (Sec.~\ref{sec:eqgrand}) -- this reasoning may be actually not always valid due to the coupling to highly pure off-resonant levels in the bath. Nevertheless, as shown by the numerical results in Sec.~\ref{sec:numres}, our theory is valid for bandwidths small enough such that the mentioned effect is not yet important, but still large enough to observe an asymptotic thermalization of the system via the relaxation process.

The initial correlation matrix $\mathcal{C}(0)$, corresponding to the assumption of equal initial occupancy of the bath levels, may be expressed as
\begin{align}
	\mathcal{C}(0)=\text{diag}[p_0,f,\ldots,f],
\end{align}
where $p_t=\mathcal{C}_{00}(t)=\langle c_0^\dagger c_0 \rangle(t)$ denotes a time-dependent occupancy of the system, and thus $p_0$ is the initial occupancy. The expression above can be rewritten as
\begin{align}
	\mathcal{C}(0)=f\mathds{1}+(p_0-f)\Lambda_0,
\end{align}
where $\mathds{1}$ is $(K+1) \times (K+1)$ identity matrix, and $\Lambda_0=\text{diag}(1,0,\ldots,0)$ with $K$ elements $0$. The time-evolved correlation matrix takes the form
\begin{align}
	\mathcal{C}(t)=f \mathds{1} + (p_0-f)\Lambda_t,
\end{align}
where $\Lambda_t=e^{i \mathcal{H}t} \Lambda_0 e^{-i \mathcal{H}t}$.

One may now note that $\Lambda_0$ corresponds to the correlation matrix of a single-particle pure state: $(\Lambda_0)_{kl} = \langle \Lambda_0|c_k^\dagger c_l |\Lambda_0\rangle$, where
\begin{align}
	|\Lambda_0\rangle = c_0^\dagger | \varnothing \rangle.
\end{align}
To recall, $| \varnothing \rangle$ denotes here a vacuum state. Correspondingly, $\Lambda_t$ is the correlation matrix of the time-evolved state $|\Lambda_t\rangle=e^{-i\hat{H}t}|\Lambda_0\rangle$. It is then known that any pure system-bath state can be transformed via a unitary matrix acting only on the bath to a Schmidt form
\begin{align}
	|\tilde{\Lambda}_t \rangle=(\alpha c_0^\dagger+\gamma \tilde{c}_1^\dagger) | \varnothing \rangle,
\end{align}
where $\tilde{c}_1^\dagger=\sum_{k=1}^K a_k c_k^\dagger$ is a certain superposition of the creation operators in the original basis, while $\alpha$ and $\gamma$ are nonnegative real numbers. A corresponding transformed correlation matrix $(\tilde{\Lambda}_t)_{kl}=\langle \tilde{\Lambda}(t)|\tilde{c}_k^\dagger \tilde{c}_l |\tilde{\Lambda}(t) \rangle$ takes the form
\begin{align}
	\tilde{\Lambda}_t=\begin{pmatrix} \alpha^2 & \alpha \gamma \\ \alpha \gamma & \gamma^2 \end{pmatrix} \oplus \text{diag}(0,\ldots,0),
\end{align}
where $\oplus$ denotes a direct sum of matrices, i.e., $A \oplus B = \text{diag}(A,B)$. The correlation matrix $\mathcal{C}(t)$ is then transformed to a form
\begin{align}
	\tilde{\mathcal{C}}(t)=f \mathds{1}+(p_0-f) \tilde{\Lambda}_t.
\end{align}
Parameters $\alpha$ and $\gamma$ can be found by using the identities $p_t=f+(p_0-f)\alpha^2$, $\tilde{\mathcal{C}}_{11}(t)=f+(p_0-f)\gamma^2$, and $p_0+f=p_t+\tilde{\mathcal{C}}_{11}(t)$; the latter identity is a consequence of the particle number conservation (or, in other words, the conservation of trace of the correlation matrix). One thus finds
\begin{align}
	\tilde{\mathcal{C}}(t)=\begin{pmatrix} p_t & \delta \\ \delta &f-\Delta_t \end{pmatrix} \oplus \text{diag}(f,\ldots,f),
\end{align}
where $\Delta_t=p_t-p_0$ and $\delta=|\sqrt{(p_0-p_t)(f-p_t)}|$. 

As one may note, after the transformation the system-bath correlation corresponds to a correlation between the system and a single mode of the transformed bath. Thus, we may focus on the reduced correlation matrix of the modes 0 and 1, denoted as $\tilde{\rho}_{(0,1)}$. Using Eq.~\eqref{denmatfromcor} it can be represented as
\begin{align} \label{denmatan}
	\tilde{\rho}_{(0,1)}=\begin{pmatrix} b_1 & 0 & 0 & 0 \\ 0 & b_2 & \delta & 0 \\ 0 & \delta & b_3 & 0 \\
		0 & 0 & 0 & b_4
	\end{pmatrix},
\end{align}
with $b_1=p_t (f-\Delta_t)-\delta^2$, $b_2=p_t(1-f+\Delta_t) + \delta^2$, $b_3=(1-p_t)(f-\Delta_t) + \delta^2$, and $b_4=(1-p_t)(1-f+\Delta_t)-\delta^2$. The partially-transposed density matrix takes then the form
\begin{align} \label{parttranspan}
	\tilde{\rho}_{(0,1)}^{T_B}=\begin{pmatrix} b_1 & 0 & 0 & \delta \\ 0 & b_2 & 0 & 0 \\ 0 & 0 & b_3 & 0 \\
		\delta & 0 & 0 & b_4
	\end{pmatrix}.
\end{align}

Finally, using Eq.~\eqref{negativity}, the entanglement negativity can be calculated as
\begin{align} \label{negan}
	\mathcal{N}=\max(0,-\lambda_1),
\end{align}
where
\begin{align}
	\lambda_1=&\frac{1}{2} \left[1-f-p_0-2f p_0 +4p_t(f+p_0-p_t) \right. \\  & \left. -\sqrt{(p_0+f-1)^2+4|(p_0-p_t)(f-p_t)|}\right] \nonumber
\end{align}
is the only eigenvalue of $\tilde{\rho}_{(0,1)}^{T_B}$ which can take negative values.

Quite notably, our method allows characterizing the system-bath entanglement using only the system observables and intensive thermodynamic parameters of the bath (specifically, the temperature and chemical potential that determine the Fermi distribution). This is generally not possible for microscopic system-bath correlations. We note that this approach can be used for the study of other types of system-bath correlations (provided that they are invariant to local unitary operations). In particular, in the Appendix~\ref{sec:mutinf} we present an analytic description of the system-bath mutual information, which was studied numerically in our previous work~\cite{ptaszynski2022}. 

\subsubsection{Analysis of the result} \label{subsec:anresults}
Let us now analyze the behavior of the entanglement negativity. We will focus on the case when the occupancy of the system undergoes a Markovian relaxation process described by the master equation~\cite{fichetti1998}
\begin{align}
	\dot{p}_t=\Gamma(f-p_t),
\end{align} 
whose solution is 
\begin{align} \label{markev}
	p_t=f+(p_0-f)e^{-\Gamma t}.
\end{align} 
We note that Markovianity of the dynamics is not a requirement of the validity of Eq.~\eqref{negan} -- it is still valid when the dynamics is non-Markovian, e.g., due to a finite bandwidth $W$. We further note that by taking the simultaneous limits $W/\Gamma \rightarrow \infty$ and $k_B T/W \rightarrow \infty$ with $(\epsilon_0-\mu)/(k_B T)=\text{const.}$ (such that the bandwidth is infinite, but occupancy of each bath level is still equal to $f$), we reach the so-called singular-coupling limit~\cite{frigerio1976, breuer2002} where both the Markovian description and our analytic theory of entanglement are exact. Thus, our approach enables us to study the system-bath entanglement in a fully Markovian regime.

\begin{figure}
	\centering
	\includegraphics[width=0.96\linewidth]{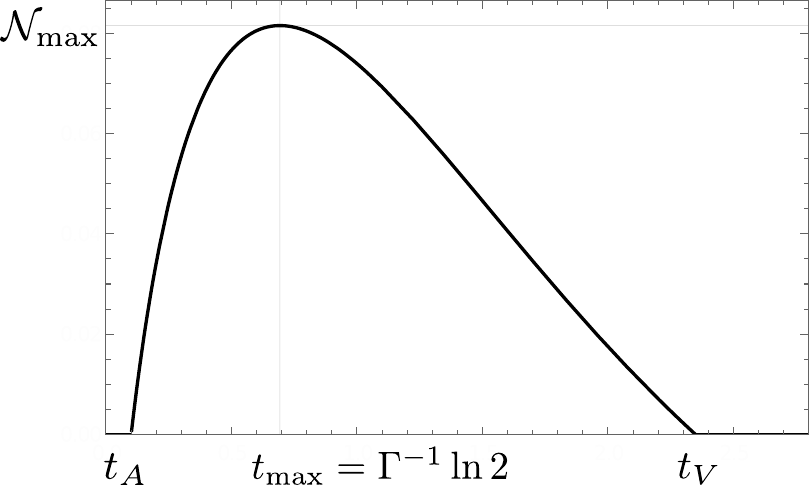}
	\caption{A scheme of the entanglement evolution during Markovian relaxation. The entanglement negativity appears at the time $t_A$, reaches a maximum value $\mathcal{N}_\text{max}$ in the moment $t_\text{max}=\Gamma^{-1} \ln 2$, and vanishes at the time $t_V$.}
	\label{fig:negschem}
\end{figure}
First of all, the theory shows that in a certain range of initial conditions $p_0$ and $f$ (which will be described later) the system-bath entanglement is generated within some interval, and exhibits a non-monotonic behavior schematically presented in Fig.~\ref{fig:negschem}. As Eq.~\eqref{negan} is not directly dependent on the coupling strength $\Gamma$ (which determines only the timescales of the entanglement evolution), this is true for an arbitrarily weak finite $\Gamma$. This conclusion will be later confirmed by numerical simulations. We consider this to be a remarkable result, as in equilibrium entanglement appears only above a certain threshold $\Gamma/(k_B T)$. This further illustrates that the applicability of Born approximation (which assumes a factorized system-bath state) for derivation of Markovian master equation cannot be naively used to infer a lack of significant system-bath correlations in the weak-coupling regime. Physically, this may be explained as follows: The presence of entanglement in our model is related to the unitary character of the microscopic global system-bath dynamics underlying the reduced description. It involves the generation of quantum coherences in the eigenbasis of the free Hamiltonian $H_S+H_E$, corresponding to off-diagonal elements of the correlation matrix. However, the reduced dynamics of the system is effectively classical, as such coherences are washed out by applying a partial trace over the state of the bath~\cite{smirne2021}.

Going into details of the entanglement behavior, we see that at the beginning of the evolution no system-bath entanglement is present until the entanglement arrival time $t_A$. This is because -- analogously to the equilibrium case -- the entanglement appears only when system-bath correlations (which gradually build up at the beginning of the evolution) reach a certain finite threshold~\cite{zyczkowski2001, yu2009}. After the time $t_A$ the entanglement negativity increases, until it reaches a maximum value $\mathcal{N}_\text{max}$ at the time $t_\text{max}$. This time corresponds to the moment when the difference between the system occupancy $p_t$ and the equilibrium population $f$ decreases to half of its initial value: $p_{t_\text{max}}-f=(p_0-f)/2$. Thus, independent on the initial parameters, it takes a universal value 
\begin{align}
	t_\text{max}=\Gamma^{-1} \ln 2,
\end{align}
which is a relaxation half-time. Finally, the entanglement vanishes at the vanishing time $t_V$. This can be explained by the phenomenon of post-thermalization (also referred to as the asymptotic factorization~\cite{cusumano2018}) analyzed in our previous paper~\cite{ptaszynski2022} (see also Appendix~\ref{sec:mutinf}): At long times the system-bath correlations gradually decrease due to reconversion into the correlations within the bath. When the correlations decrease below a certain finite threshold, the state becomes separable, which is sometimes called as an entanglement sudden death~\cite{zyczkowski2001, yu2009}. We further note that (for the considered Markovian relaxation) the entanglement arrival and vanishing times are not independent, but related via the equation
\begin{align}
	e^{-\Gamma t_A}=1-e^{-\Gamma t_V}.
\end{align}
Furthermore, in general the entanglement negativity obeys a symmetry relation
\begin{align}
	\mathcal{N}(t_1)=\mathcal{N}(t_2) \quad \text{for} \quad e^{-\Gamma t_1}=1-e^{-\Gamma t_2}.
\end{align}

\begin{figure}
	\centering
	\includegraphics[width=0.95\linewidth]{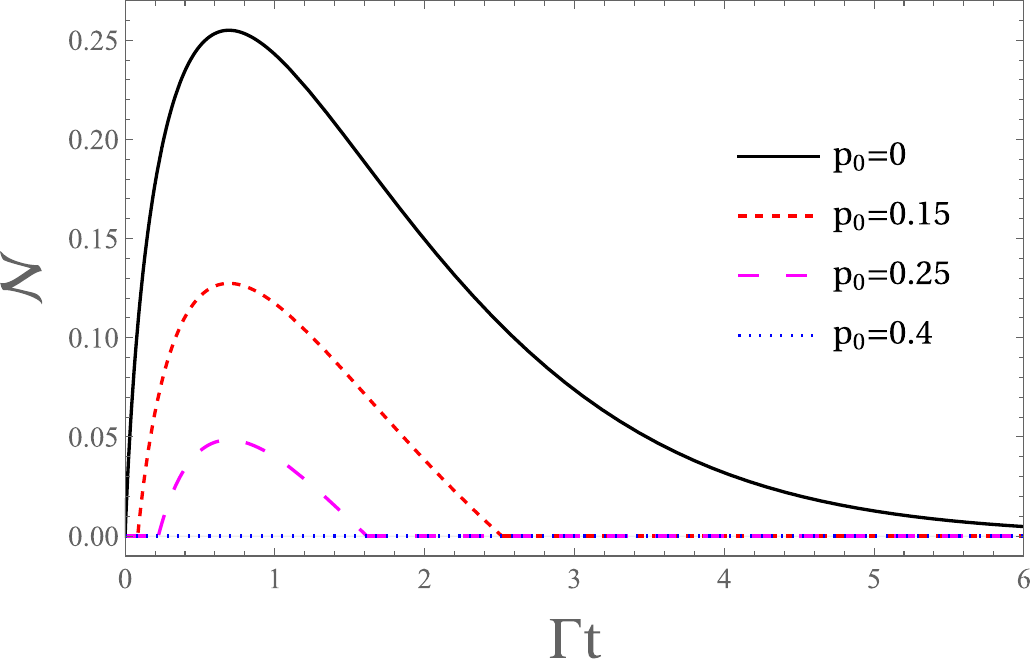}
	\caption{Entanglement negativity as a function of time for different initial system occupancies $p_0$. Results for $\mu=\epsilon_0+k_B T$.}
	\label{fig:ent-an}
\end{figure}
A quantitative analysis of the entanglement negativity shows that its behavior strongly depends on the initial occupancy $p_0$. This is presented in Fig.~\ref{fig:ent-an}. The results are plotted for $f=e/(1+e)$, which corresponds to $\mu=\epsilon_0+k_B T$. We first note that for an initial pure state $p_0=0$ (black solid line) the entanglement is created immediately ($t_A=0$) and goes asymptotically to zero only for the infinite time $t_V \rightarrow \infty$; the analogous results are obtained for $p=1$ (not shown). We note that an immediate generation of the system-bath entanglement, for a system initialized in a pure state, was previously shown for the bosonic case~\cite{eisert2002}; however, the infinite vanishing time is a peculiar feature of the considered model, as it is not observed for bosons~\cite{hilt2009}.

For initial mixed states with $p_0=0.15$ (red dashed line) or $p_0=0.25$ (violet large dahed line), the entanglement is smaller than for a pure state. The entanglement arrival and vanishing times $t_A$ and $t_V$ usually need to be determined numerically. However, approximate analytical expressions may be derived by considering the regime of a high initial purity ($p\approx 0$ or $p_0 \approx 1$). This is done by expanding $\lambda_1$ as the power series of $t$ and $p_0$ or $1-p_0$, and then finding $t_A$ by solving $\lambda_1=0$ for the lowest orders of the expansion. For $p_0 \approx 0$ one finds
\begin{align}
	t_A &\approx \Gamma^{-1} \frac{p_0(1-f)}{f}, \\
	t_V &\approx -\Gamma^{-1} \ln \left[\frac{p_0(1-f)}{f} \right],
\end{align}
while for $p_0 \approx 1$
\begin{align}
	t_A &\approx \Gamma^{-1} \frac{(1-p_0)f}{1-f}, \\
	t_V &\approx -\Gamma^{-1} \ln \left[\frac{(1-p_0)f}{1-f} \right].
\end{align}
As these expressions show, for a high initial purity entanglement appears almost immediately and vanishes for times orders of magnitude longer than the relaxation time.

\begin{figure}
	\centering
	\includegraphics[width=0.8\linewidth]{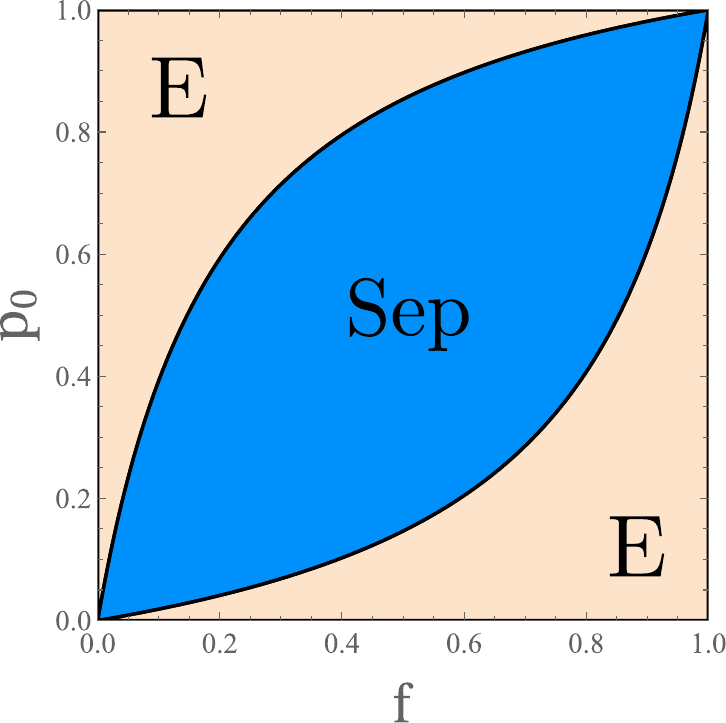}
	\caption{The entanglement phase diagram showing the ranges of initial parameters $f$ and $p_0$ for which the system develops a transient system-bath entanglement (the orange regions denoted ``E'') or not (the blue region denoted ``Sep'').}
	\label{fig:phasediag}
\end{figure}
Finally, for a highly mixed state (here $p_0=0.4$) entanglement does not appear at all (blue dots). This is graphically presented in the entanglement phase diagram (Fig.~\ref{fig:phasediag}), where the range of initial conditions, for which the entanglement is not generated, corresponds to a lemon-shaped region in the middle of the graph. Interestingly, we note that (in a certain range of $p_0$) the entanglement appears even for $f=1/2$, which corresponds to a maximally mixed state of the bath. Previously, entanglement with maximally mixed baths has been shown to be impossible for qubits undergoing pure dephasing~\cite{roszak2015}, while it is possible for higher dimensions of the Hilbert space of the system~\cite{roszak2018} or a non-purely dephasing evolution~\cite{roszak2015, salamon2022}.

\subsection{Numerical results} \label{sec:numres}

\subsubsection{Entanglement negativity $\mathcal{N}_M$ for different cutoffs $M$} \label{subsec:cutoff}
\begin{figure}
	\centering
	\includegraphics[width=0.9\linewidth]{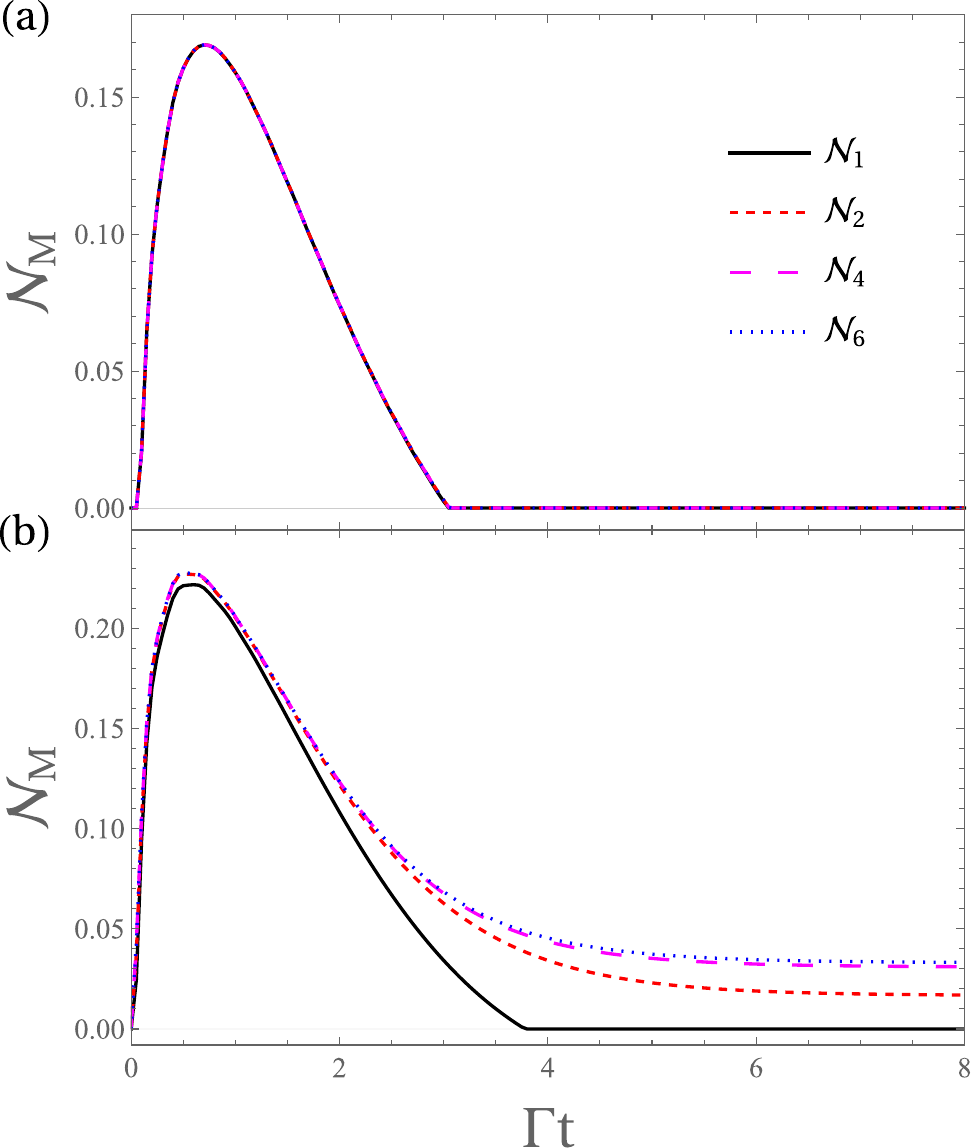}
	\caption{Evolution of the entanglement negativities $\mathcal{N}_M$ for different cutoffs $M$ with $\Gamma=0.01 k_B T$ (a) and $\Gamma=0.5 k_B T$ (b). Parameters: $p_0=0.1$, $\epsilon_0=0$, $\mu=k_B T$, $W=50 \Gamma$ and $K=400$.}
	\label{fig:compM}
\end{figure}

Let us now present the numerical results obtained using the methods described in Secs.~\ref{subsec:cormat} and~\ref{subsec:nunnegcalc}. First, we compare the evolution of partial negativities $\mathcal{N}_M$ for different cutoffs $M$. We consider the case of a weak ($\Gamma=0.01 k_B T$) and a strong ($\Gamma=0.5 k_B T$) system-bath coupling. The results are presented in Fig.~\ref{fig:compM}. We note that for a weak coupling the curves approximately coincide. This is because, as shown by the analytic theory, in the weak-coupling regime the entanglement is concentrated in correlations between a system and a single mode of the transformed bath state. For a strong coupling, the calculated negativities approximately coincide at short times but start to deviate for long times. In particular, $\mathcal{N}_1$ vanishes at a certain moment, whereas partial negatives $\mathcal{N}_M$ do not disappear for $M>1$. As shown later, this is because the entanglement converges to a finite value predicted by the equilibrium theory from Sec.~\ref{sec:eqgrand}. Analogously to the equilibrium entanglement, the asymptotic long-time value of $\mathcal{N}_M$ depends on $M$. However, $\mathcal{N}_4$ and $\mathcal{N}_6$ are very close to each other, which suggests that (for the parameters considered) $\mathcal{N}_6$ is a good approximation of the total negativity $\mathcal{N}$.

\subsubsection{Finite size effects -- Poincar\'{e} recurrences}
\begin{figure}
	\centering
	\includegraphics[width=0.94\linewidth]{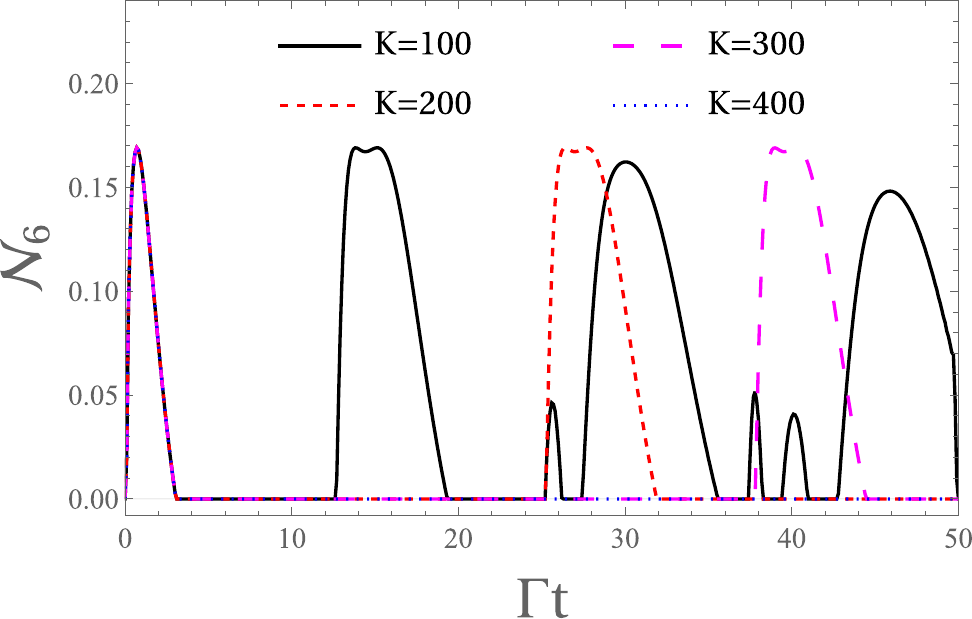}
	\caption{Entanglement negativity $\mathcal{N}_6$ as a function of time for different numbers of bath levels $K$. Results for $\Gamma=0.01k_B T$ and other parameters as in Fig.~\ref{fig:compM}.}
	\label{fig:poincare}
\end{figure}
The derivation of the master equation assumes the infinite bath limit $K \rightarrow \infty$~\cite{schaller2014}, while using the correlation matrix approach we simulate baths with finite sizes. Let us now consider the effect of a finite bath size. In Fig.~\ref{fig:poincare} we present the evolution of the entanglement negativity for different numbers of bath levels $K$, while keeping a fixed bandwidth; we thus change the separation of energy levels in the bath. As one may note, we consider much longer time scales than previously presented. For short times, entanglement is approximately independent of the bath size. This differs, e.g., from the pure dephasing of a qubit attached to a harmonic oscillator bath, where entanglement decreases with the bath size and vanishes for infinite baths~\cite{salamon2017}. For longer times we observe sudden revivals of entanglement at times proportional to the bath size. Such sudden death and rebirth dynamics of entanglement is characteristic for mixed states undergoing a unitary evolution~\cite{bartkowiak2011}. In our model, the observed revivals are related to Poincar\'{e} recurrences -- periodic returns of a finite system undergoing a unitary dynamics to a proximity of its initial state~\cite{pucci2013}. Indeed, the entanglement revival time corresponds to the Poincar\'{e} recurrence time $t_P=2 \pi/\Delta \epsilon=2\pi K/W$, where $\Delta \epsilon$ is the distance between the bath levels.

\subsubsection{Finite $\Gamma$} \label{sec:finitegamma}
\begin{figure}
	\centering
	\includegraphics[width=0.9\linewidth]{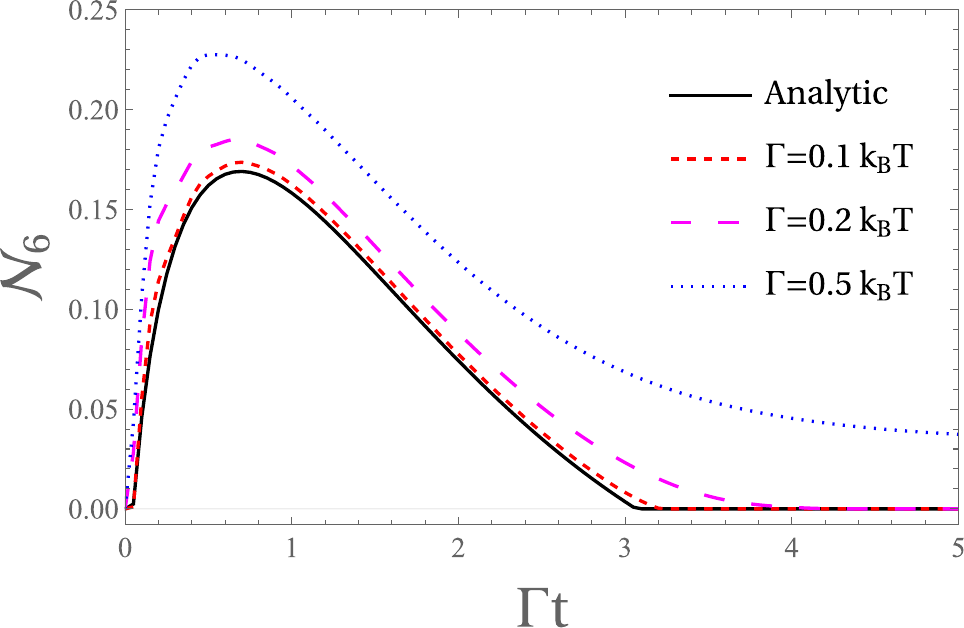}
	\caption{Entanglement negativity $\mathcal{N}_6$ as a function of time for different values of the coupling strength $\Gamma$, compared with the analytic formula for the Markovian dynamics. Other parameters as in Fig.~\ref{fig:compM}.}
	\label{fig:compG}
\end{figure}
\begin{figure}
	\centering
	\includegraphics[width=0.9\linewidth]{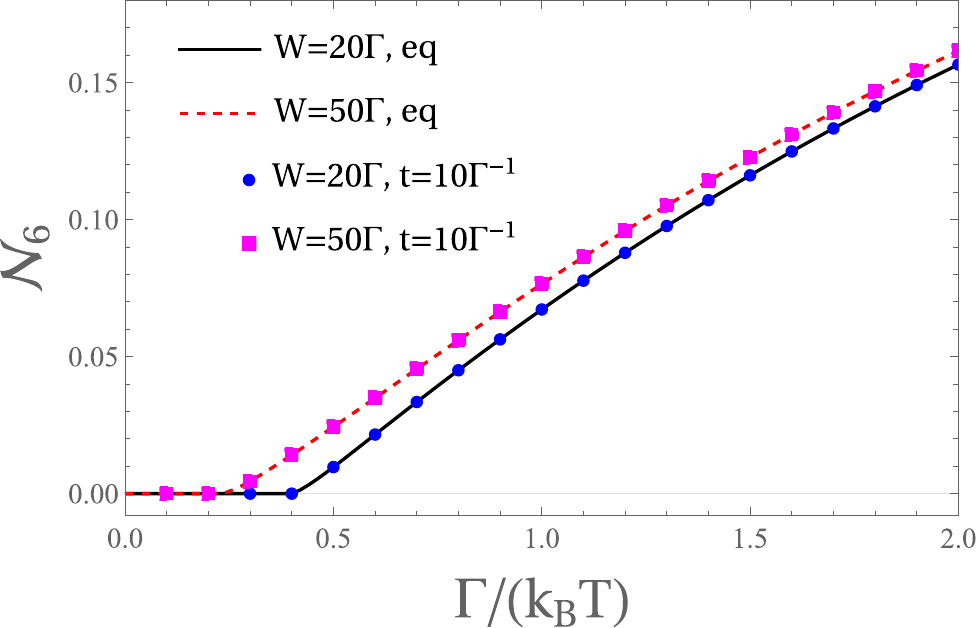}
	\caption{The equilibrium entanglement negativity $\mathcal{N}_6$ (denoted ``eq'', lines) compared with the entanglement generated during the transient evolution for $t=10 \Gamma^{-1}$ (points). The considered bandwidths denoted in the graph. Other parameters as in Fig.~\ref{fig:compM}.}
	\label{fig:compeqtrans}
\end{figure}

Let us now analyze in detail the role of the coupling strength $\Gamma$.  The results are presented in Fig.~\ref{fig:compG}. As one can observe, for a weak $\Gamma=0.1 k_B T$ the calculated negativity agrees well with the predictions of the Markovian theory. Indeed, to the lowest order of $\Gamma$, the entanglement magnitude does not depend on the coupling strength to the bath. This is consistent with the fact that within the analytic theory $\Gamma$ determines only the timescales of entanglement evolution, but not its magnitude. For a stronger coupling $\Gamma=0.2 k_B T$ the analytic theory underestimates the entanglement; however, the qualitative behavior of its evolution is still similar. Finally, as already shown in Fig.~\ref{fig:compM}, for a very strong coupling $(\Gamma=0.5 k_B T$) entanglement does not vanish at long times, but rather saturates at some finite value. As further shown in Fig.~\ref{fig:compeqtrans}, this long-time asymptotic value of entanglement (here calculated for $t=10\Gamma^{-1}$) perfectly agrees with the equilibrium entanglement calculated for the same parameters. This is a remarkable result, as the long-time convergence to equilibrium is not a trivial issue for open quantum systems strongly coupled to the bath, even if one considers just a reduced state of a system~\cite{trushechkin2022}. Indeed, as we will later show, this no longer holds true in the presence of strongly non-Markovian effects that suppress thermalization.

\begin{figure}
	\centering
	\includegraphics[width=0.9\linewidth]{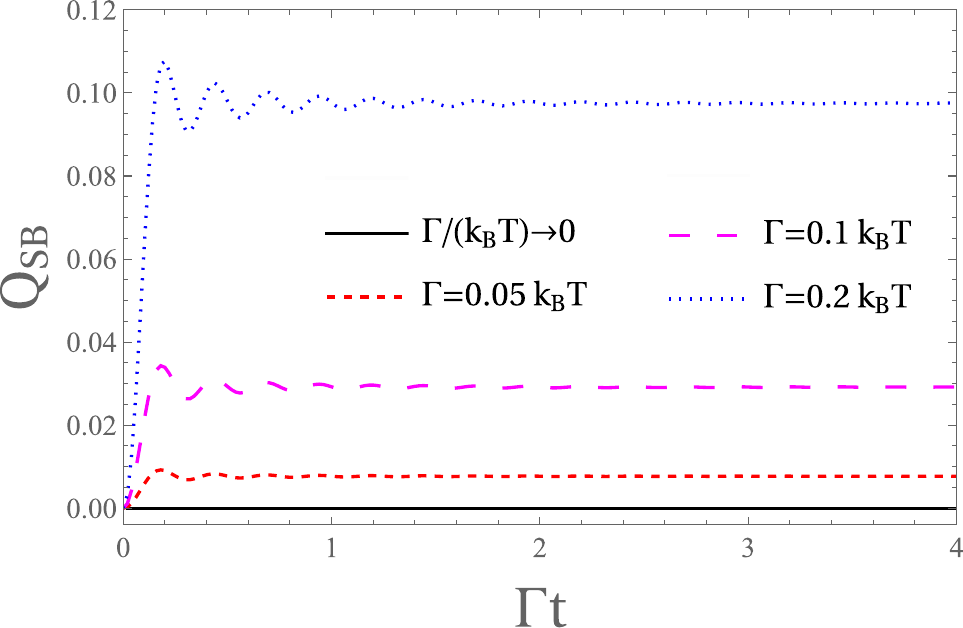}
	\caption{The heat asymmetry $Q_{SB}$ as a function of time for different values of the coupling strength $\Gamma$. Other parameters as in Fig.~\ref{fig:compM}.}
	\label{fig:heatasym}
\end{figure}

Our approach enables us to investigate also the alleged link between entanglement and strong-coupling thermodynamic effects reported in Ref.~\cite{bernardo2021}. This study considered a setup in which both the system and the bath consisted of a single qubit. It was observed that during the transient evolution the entanglement negativity is approximately proportional to the heat asymmetry defined as 
\begin{align}
	Q_{SB}=\Delta E_S+\Delta E_B,
\end{align}
where $\Delta E_\alpha=E_\alpha(t)-E_\alpha(0)$ ($\alpha \in \{S,B\}$) is the energy change of the system or the bath, and $E_\alpha(t)$ is the energy at the time $t$. This quantity is related to system-bath interaction energy, and thus vanishes in the weak-coupling Markovian regime when $\Delta E_S=-\Delta E_B$~\cite{breuer2002, benenti2017}. The observed proportionality of the entanglement negativity and $Q_{SB}$ led the author of Ref.~\cite{bernardo2021} to the conclusion that the presence of heat asymmetry is responsible for the generation of the system-bath entanglement.

Here we analyze the heat asymmetry in our model. The energy of the system is calculated as $E_S(t)=\epsilon_0 p_t$, while the bath energy as $E_B(t)=\sum_{k=1}^K \epsilon_k \mathcal{C}_{kk}(t)$. The results for different coupling strengths $\Gamma$ are presented in Fig.~\ref{fig:heatasym}. As one can observe, the heat asymmetry, after initial transient oscillations, saturates at some finite value. Therefore, its evolution is qualitatively very different from the non-monotonic behavior of the entanglement negativity. Furthermore, the value of the heat asymmetry exhibits a strong (supralinear) dependence on the coupling strength, while the entanglement negativity exhibits no such strong dependence. In particular, entanglement may also be generated in the limit of $\Gamma/(k_B T) \rightarrow 0$ (for which the analytic theory is exact), when there is no heat asymmetry. Thus, the relation between the entanglement negativity and the heat asymmetry observed in Ref.~\cite{bernardo2021} appears to be a specific feature of the considered model rather than a generic rule. 

\subsubsection{Finite bandwidth}
\begin{figure}
	\centering
	\includegraphics[width=0.9\linewidth]{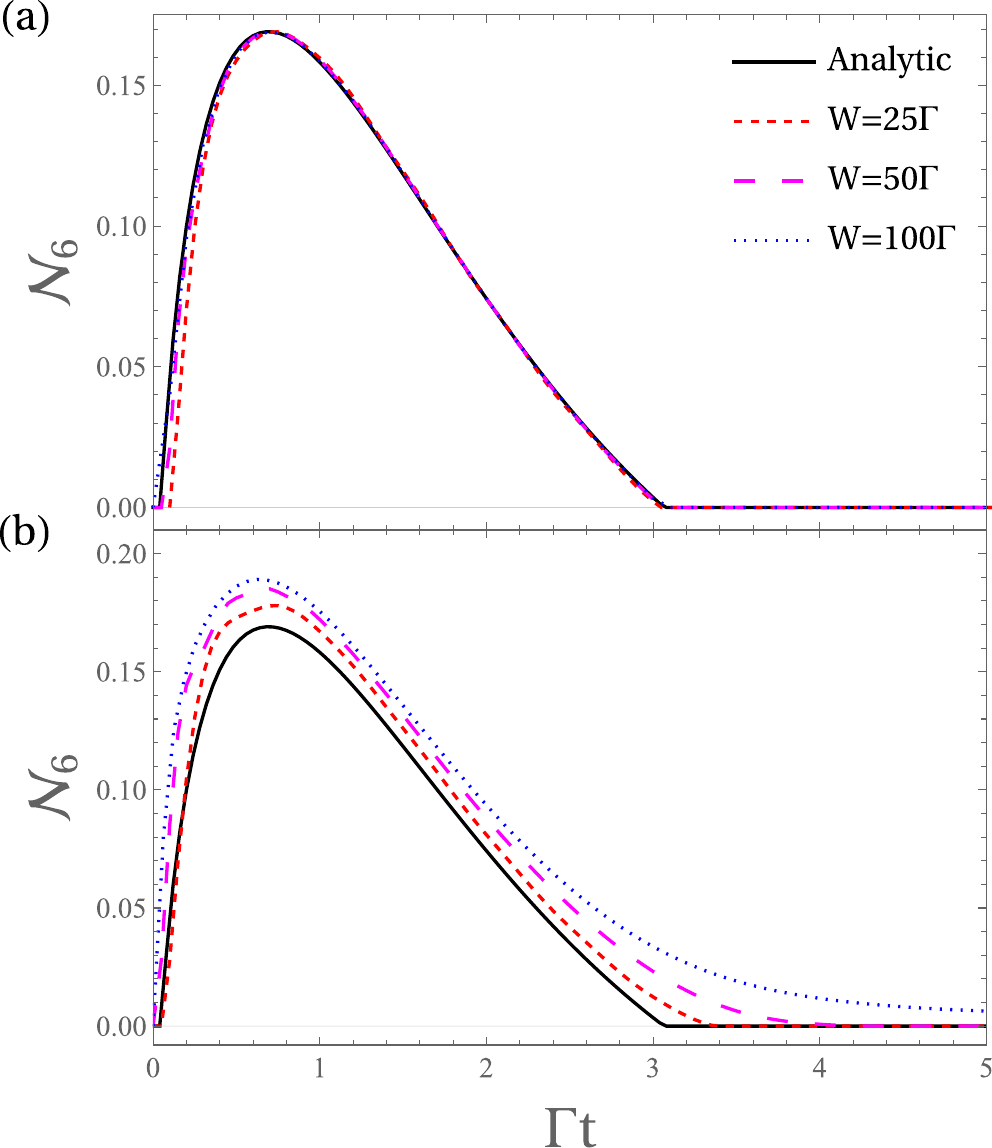}
	\caption{Entanglement negativity as a function of time for different bandwidths $W$ with (a) $\Gamma=0.01k_B T$ and (b) $\Gamma=0.2 k_B T$, compared with the analytic formula for the Markovian dynamics. Results for $K=8W/\Gamma$ and other parameters as in Fig.~\ref{fig:compM}.}
	\label{fig:bandwidth-neg}
\end{figure}
\begin{figure}
	\centering
	\includegraphics[width=0.9\linewidth]{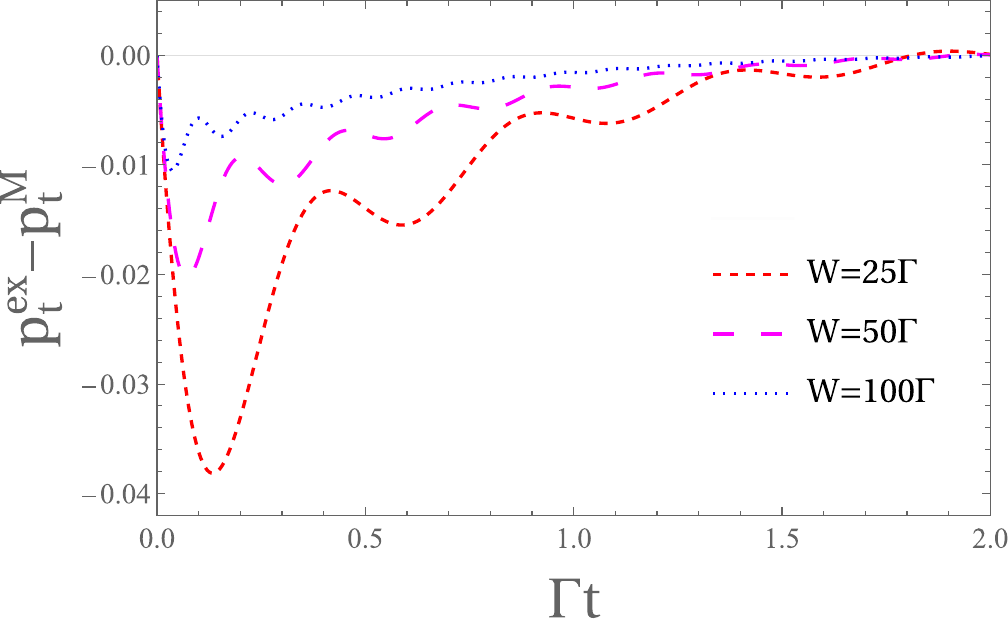}
	\caption{Difference of the system occupancy $p_t$ between the exact ($p_t^\text{ex}$) and the Markovian ($p_t^\text{M}$) dynamics for $\Gamma=0.01 k_B T$ and different bandwidths $W$. Parameters as in Fig.~\ref{fig:bandwidth-neg}.}
	\label{fig:bandwidth-oc}
\end{figure}

Let us now consider the influence of the finite bandwidth $W$. To keep the distance between bath levels fixed, the bath size $K$ is taken to be proportional to $W$. The evolution of entanglement negativity for a weak ($\Gamma=0.01k_B T$) and a relatively strong ($\Gamma=0.2k_B T$) coupling strength, compared with the analytic theory for the Markovian case, is presented in Fig.~\ref{fig:bandwidth-neg}. As one can observe, for a weak coupling the Markovian approach overestimates the entanglement negativity at short times for small bandwidths (here $W=25 \Gamma$), as well as underestimates the entanglement arrival time $t_A$. However, for a large bandwidth $W=100 \Gamma$ the difference becomes negligible. The observed behavior is the result of the deviation of the system occupancy $p_t$ from the predictions of the Markovian master equation (see Fig.~\ref{fig:bandwidth-oc}) . It is notable mainly at times shorter than the relaxation time $\Gamma^{-1}$. Its magnitude is inversely proportional to the bandwidth, as the master equation becomes exact in the infinite bandwidth limit.

In Fig.~\ref{fig:bandwidth-neg}~(b) we present the case of a stronger coupling $\Gamma=0.2 k_B T$. In this case, the analytic theory underestimates the entanglement negativity, which -- analogously the equilibrium case -- increases with the bandwidth. The same behavior is observed for the entanglement vanishing time $t_V$. For a very large bandwidth $\Gamma=100\Gamma$ the entanglement does not vanish at all, but rather saturates at some finite value, which is consistent with the equilibrium predictions (cf.~Fig.~\ref{fig:tth}). We expect that such behavior is also present for smaller $\Gamma$, albeit beyond the range of bandwidths that we can simulate.

\subsubsection{Role of bound states} \label{sec:bound}
In Sec.~\ref{sec:finitegamma} it was shown that for long times the entanglement negativity approaches the value predicted by the global Gibbs state. However, this only holds when the dynamics is thermalizing, i.e., when the system approaches the equilibrium state independent of the initial conditions. Thermalization can be suppressed, e.g., by the presence of bound states, i.e., eigenstates of the single-particle Hamiltonian $\mathcal{H}$ strongly localized in the system~\cite{cai2014, xiong2015, yang2015, jussiau2019}. As shown in our previous work~\cite{ptaszynski2022}, this can suppress the decay of the system-bath mutual information, which quantifies both quantum and classical correlations. Here we demonstrate that this is true also for entanglement.

\begin{figure}
	\centering
	\includegraphics[width=0.9\linewidth]{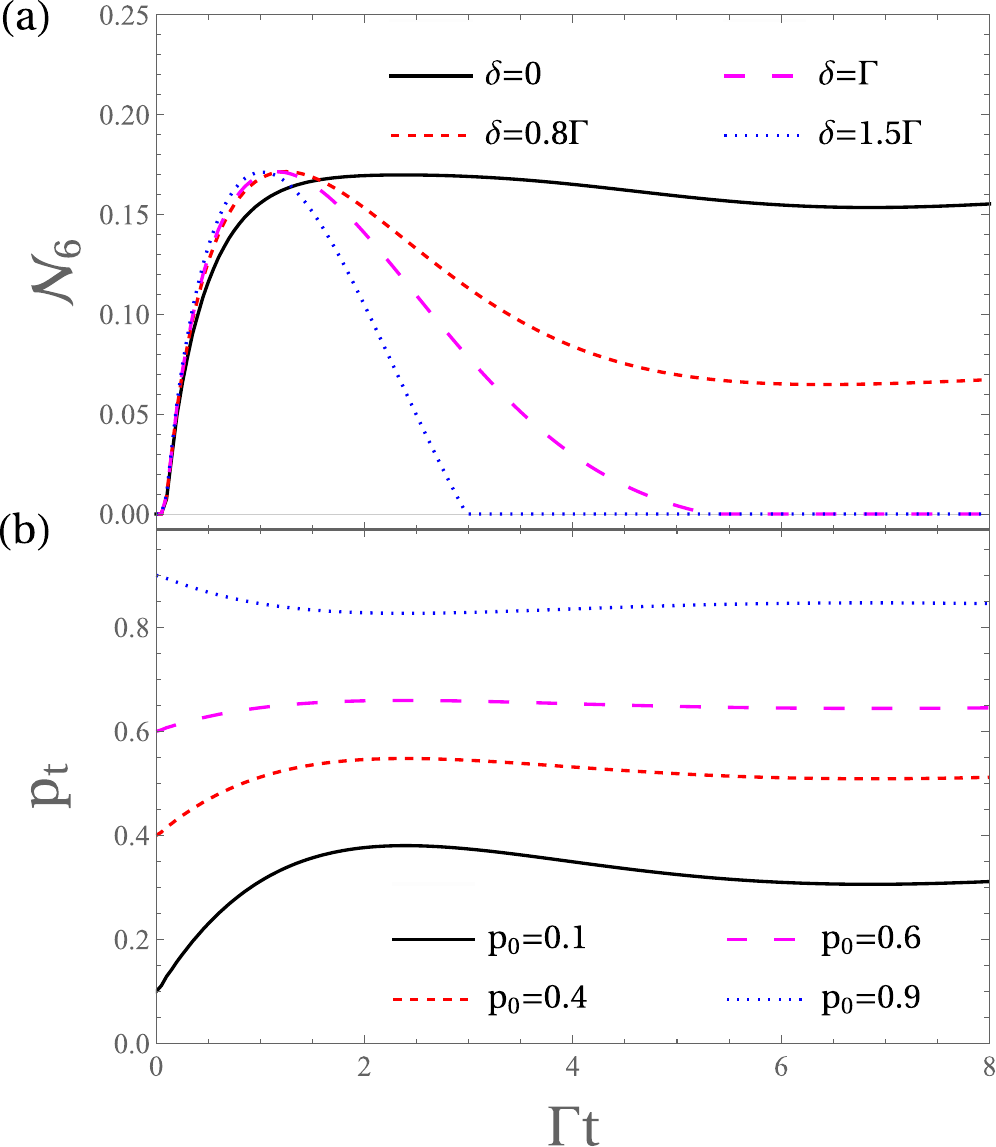}
	\caption{(a) Entanglement negativity $\mathcal{N}_6$ as a function of time for different distances of the system energy  $\epsilon_0$ from the band edge, with $p_0=0.1$. (b) Evolution of the system occupancy $p_t$ for different initial occupancies $p_0$, with $\delta=0$. Results for $\epsilon_0=W/2-\delta$, $\mu=\epsilon_0+k_B T$, $W=50 \Gamma$, and $K=400$.}
	\label{fig:bound}
\end{figure}
In the analyzed model, the bound state is generated when the energy level of the system $\epsilon_0$ is placed close to the band edge $\omega=W/2$, where the spectral density of the bath $\Gamma(\omega)$ drops from $\Gamma$ to 0~\cite{jussiau2019}. We parameterize the distance from the band edge as $\delta =W/2-\epsilon_0$. The entanglement behavior for different values of $\delta$ is presented in Fig.~\ref{fig:bound}~(a). As one may observe, for a sufficiently high value of displacement ($\delta=1.5 \Gamma$) entanglement undergoes the ``sudden death'' at time $t_V \approx 3 \Gamma^{-1}$. For a lower displacement value ($\delta=\Gamma$) entanglement is preserved for longer times, but ultimately still vanishes. Finally, when the system energy is very close to the band edge ($\delta \lessapprox 0.8 \Gamma$), entanglement is preserved also in the long time limit. 

We underline that, in contrast to the strong-coupling case ($\Gamma=0.5 k_B T$) presented in Fig.~\ref{fig:compG}, the preservation of entanglement at long times is here not a result of the convergence to the equilibrium value predicted by the global Gibbs state. In fact, for the parameters considered here there is no equilibrium entanglement. Instead, the preservation of entanglement is a consequence of the suppression of thermalization, namely, the fact that the long-time occupancy of the system does not converge to equilibrium, but rather depends on the initial state [see Fig.~\ref{fig:bound}~(b)]. Indeed, the entanglement negativity can be described by the analytic weak-coupling theory [Eq.~\eqref{negan}], with $p_t$ given by the actual occupancy rather than predictions of the Markovian master equation. This result shows that while non-Markovian effects are not essential for the generation of the system-bath entanglement, they may lead to its long time preservation. We note that an analogous suppression of the entanglement decay by non-Markovian effects was previously explored in Refs.~\cite{maniscalo2008, bellomo2008} in the context of entanglement within an open quantum system.

\section{Voltage-driven junction} \label{sec:volt}
\subsection{Model} \label{sec:voltmod}
Finally, let us consider the case where the energy level of the system is connected to two fermionic baths $\alpha \in \{L,R\}$ with the same temperature $\beta$, but different chemical potentials $\mu_L$ and $\mu_R$. The open system is described by a generalized version of the Hamiltonian~\eqref{hamnrl},
		\begin{align} \label{hamnrl2b} \nonumber
	\hat{H}=
	&\epsilon_{0} c^\dagger_{0} c_{0} +\sum_{\alpha \in \{L,R\}} \sum_{k=1}^{K} \epsilon_{\alpha k} c_{\alpha k}^\dagger c_{\alpha k} \\ & + \sum_{\alpha \in \{L,R\}} \sum_{k=1}^{K} \left( t_{\alpha k} c^\dagger_{0} c_{\alpha k} + \text{h.c.} \right).
\end{align}
As in the case of a single bath, the energy levels $\epsilon_{\alpha k}$ are uniformly distributed throughout the interval $[-W/2,W/2]$ and the tunnel couplings are parameterized as $\Gamma_\alpha=2 \pi t_{\alpha k}^2 (K-1)/W$. The chemical potentials are parameterized as $\mu_L=\bar{\mu}+V/2$ and $\mu_R=\bar{\mu}-V/2$, where $\bar{\mu}$ and $V$ are the average chemical potential and $V$ is the voltage bias, respectively. Similarly, the coupling strengths are parameterized as $\Gamma_L=(1+a)\Gamma$ and $\Gamma_R=(1-a)\Gamma$, where $a$ is the asymmetry coefficient. The initial correlation matrix is defined as \begin{align} \nonumber
\mathcal{C}(0)=\text{diag}[&p_0,f_L(\epsilon_{L1}),\ldots,f_L(\epsilon_{LK}), \\ &f_R(\epsilon_{R1}),\ldots,f_R(\epsilon_{RK})],
\end{align}
where $p_0$ is the initial occupancy of the system and $f_\alpha(\epsilon)=\{ 1+\exp [\beta (\epsilon-\mu_\alpha)]\}^{-1}$ is the Fermi distribution of the bath $\alpha$.

\subsection{Transient dynamics} \label{subsec:volttrans}
\begin{figure}
	\centering
	\includegraphics[width=0.9\linewidth]{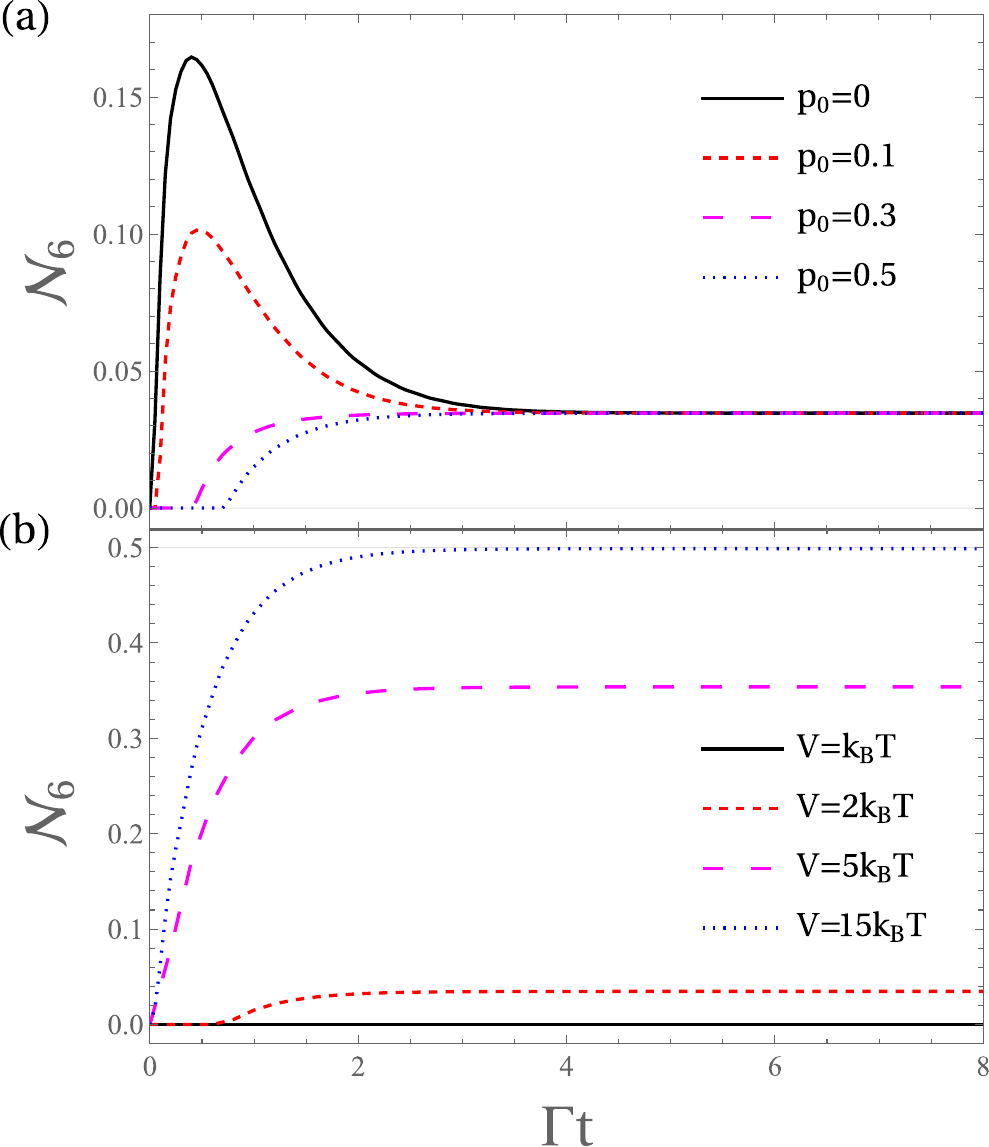}
	\caption{Entanglement negativity $\mathcal{N}_6$ as a function of time for different initial states of the system (a) and applied voltages (b). Results for $p_0=0.5$, $\Gamma=0.01 k_B T$, $a=0$, $\epsilon_0=\bar{\mu}=0$, $V=2 k_B T$, $W=50 \Gamma$, and $K=300$, unless denoted otherwise in the graph.}
	\label{fig:transient}
\end{figure}
In the first step, we analyze a transient dynamics of the voltage-driven junction. In Fig.~\ref{fig:transient}~(a) we show the behavior of entanglement for a moderately high voltage $V=2 k_B T$ and a weak coupling strength $\Gamma=0.01 k_B T$, with different initial occupancies of the system $p_0$. As can be observed, the short-time dynamics of entanglement depends on the purity of the initial state: for high purity [$p_0$ close to 0] entanglement is nearly immediately generated and reaches a maximum value for times comparable to the relaxation time $\Gamma^{-1}$. In contrast, for low purities [$p_0$ close to 1/2] entanglement is generated after a longer time. Nevertheless, for all initial states the entanglement negativity reaches the same finite asymptotic value at long times. This occurs also for a weak coupling strength $\Gamma=0.01 k_B T$, for which entanglement is not present at equilibrium. This demonstrates that nonequilibrium driving may lead to long-time preservation of the system-bath entanglement even for a weak coupling to the bath. Indeed, such a conclusion can already be drawn from the results presented in Ref.~\cite{sharma2015}, where the authors observed the mutual information between the system and the baths exceeding the value $\ln \text{dim} \mathcal{H}_S =\ln 2$ (where $\text{dim} \mathcal{H}_S$ is the dimension of the Hilbert space of the system), which is a maximum value of the mutual information for separable states~\cite{cerf1997, vollbrecht2002}.

As further shown in Fig.~\ref{fig:transient}~(b), both the short time dynamics of entanglement and its asymptotic value strongly depend on the voltage. For high voltages $V \gtrapprox 5 k_B T$, entanglement is created almost immediately and reaches larger asymptotic values. In particular, for a very high voltage $V=15 k_B T$ the asymptotic entanglement negativity is close to the maximum value 1/2. For lower voltages (here $V=2 k_B T$) entanglement is formed after a longer time and saturates at lower values. Finally, for a very small voltage $V=k_B T$ entanglement is not created at all. This voltage dependence will be the main focus of the later analysis of the steady-state entanglement.

\begin{figure}
	\centering
	\includegraphics[width=0.9\linewidth]{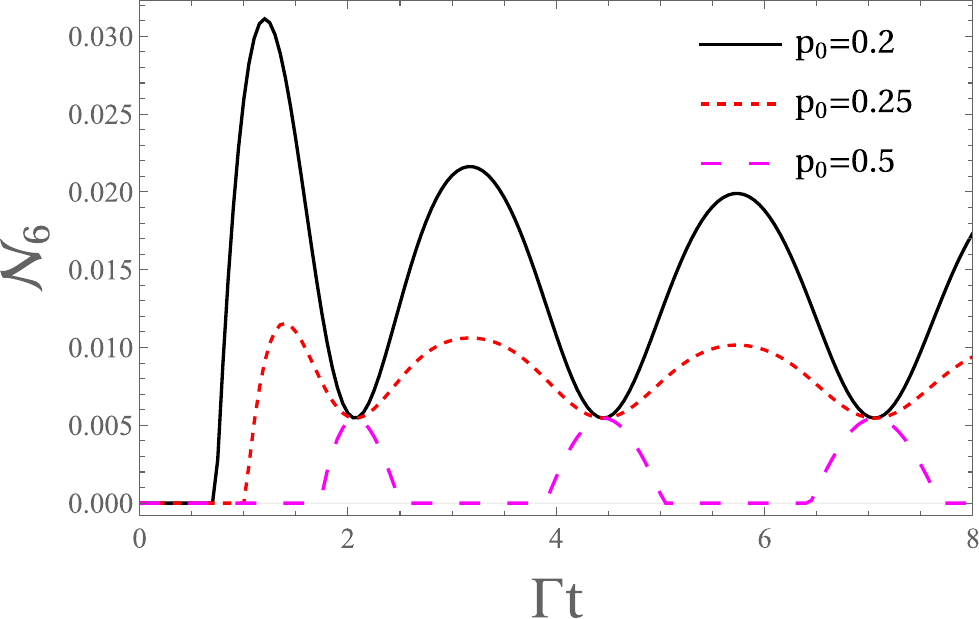}
	\caption{Entanglement negativity $\mathcal{N}_6$ as a function of time for small bandwidth $W=2\Gamma$ and different initial states of the system. Results for $\Gamma=0.01 k_B T$, $a=0$, $\epsilon_0=\bar{\mu}=0$, $V=1.8 k_B T$, and $K=200$.}
	\label{fig:transient-band}
\end{figure}
We note that the convergence to a steady state independent of the initial state is observed only when the bandwidth is sufficiently large ($W \gtrapprox 3\Gamma$). For a smaller bandwidth, as in the single-bath case considered in Sec.~\ref{sec:bound}, this no longer holds due to the presence of the bound states. This is illustrated in Fig.~\ref{fig:transient-band}. As shown, for a small bandwidth $W=2\Gamma$ the entanglement negativity depends on the initial state also for long times, and exhibits oscillations that apparently survive for arbitrarily long times. Such ``eternal oscillations'' are characteristic for the bound states~\cite{xiong2015,jussiau2019, yang2015, ali2017}. Furthermore, for the initial fully mixed state ($p_0=0.5$) one can observe periodic deaths and revivals of entanglement. Such a behavior is typical for the entanglement dynamics in non-Markovian systems~\cite{bellomo2007, mazzola2009}. Interestingly, one can observe that at certain moments entanglement reaches the same value, independent of the initial state, but later again diverges; however, this is true only when the energy level of the system is placed in the center of the band ($\epsilon_0=0$), and thus is a result of the model symmetry,

\subsection{Steady state -- analytic theory}
\subsubsection{Derivation} \label{subsec:ander}
As in the case of transient dynamics (see Sec.~\ref{subsec:andert}), the steady-state entanglement in the weak-coupling regime ($\Gamma \ll k_B T$) can be described using an analytic theory. Analogously to the previous case, it is applicable when the bandwidth is neither too small (such that the system reaches a steady state independent of the initial conditions) nor too large (such that coupling to highly pure off-resonant levels is not yet important). Based on the same arguments, we take the level occupancies of both reservoirs to be energy-independent and equal to $f_L=f_L(\epsilon_0)$ and $f_R=f_R(\epsilon_0)$. To denote the levels in the bath, we now reexpress the indexes as ${Lk}=k$ and $Rk=K+k$. We then consider the evolution of the correlation matrix from the initial uncorrelated state to the steady state. As the steady-state entanglement is independent of the initial state of a system, without loss of generality we fix the initial occupancy as $p_0=f_R$. The initial correlation matrix $\mathcal{C}(0)$ may then be expressed as
\begin{align}
	\mathcal{C}(0)=\text{diag}[f_R,f_L,\ldots,f_L,f_R,\ldots,f_R].
\end{align}
The expression above can be rewritten as
\begin{align}
	\mathcal{C}(0)=f_R \mathds{1}+(f_L-f_R)\Lambda_0,
\end{align}
where $\mathds{1}$ is a $(2K+1) \times (2K+1)$ identity matrix, and the matrix $\Lambda_0=\text{diag}(0,1,\ldots,1,0,\ldots,0)$ contains $K$ elements $1$ at positions $1,\ldots,K$. 

We now use the same approach as in Sec.~\ref{subsec:andert} by noting that $\Lambda_0$ corresponds to the correlation matrix of a pure state
\begin{align}
	|\Lambda_0\rangle = c_K^\dagger \ldots c_1^\dagger | \varnothing \rangle.
\end{align}
Following the same steps as before, the transformed correlation matrix takes the form
\begin{align}
	\tilde{\mathcal{C}}(t)=f_R \mathds{1}+(f_L-f_R) \tilde{\Lambda}_t,
\end{align}
where $\tilde{\Lambda}_t$ is expressed as
\begin{align}
	\tilde{\Lambda}_t=\begin{pmatrix} \alpha^2 & \alpha \gamma \\ \alpha \gamma & \gamma^2 \end{pmatrix} \oplus \text{diag}(1,\ldots,1,0,\ldots,0),
\end{align}
with $K-1$ elements $1$. The parameters $\alpha$ and $\gamma$ can be found using the identities $\tilde{\mathcal{C}}_{00}(t)=f_R+(f_L-f_R)\alpha^2$ and $\tilde{\mathcal{C}}_{11}(t)=f_R+(f_L-f_R)\gamma^2$. We further focus on long times, when $\tilde{\mathcal{C}}_{00}(t)$ is equal to the stationary occupancy of the system $p_\text{st}$, and we require $\tilde{\mathcal{C}}_{11}(t)+p_\text{st}=f_L+f_R$ due to the particle number conservation. One thus finds a long-time asymptotic form of the correlation matrix (which corresponds to the steady state)
\begin{align}
	\tilde{\mathcal{C}}_\text{st}=\begin{pmatrix} p_\text{st} & \delta \\ \delta &\tilde{\mathcal{C}}_{11}^\text{st} \end{pmatrix} \oplus \text{diag}(f_L,\ldots,f_L,f_R,\ldots,f_R),
\end{align}
where $\tilde{\mathcal{C}}_{11}^\text{st}=f_L+f_R-p_\text{st}$ and $\delta=|\sqrt{(f_L-p_\text{st})(f_R-p_\text{st})}|$. The partially transposed matrix of modes 0 and 1 is then given by Eq.~\eqref{parttranspan} with $\delta$ as above, $b_1=p_\text{st} \tilde{\mathcal{C}}_{11}^\text{st}-\delta^2$, $b_2=p_\text{st} (1-\tilde{\mathcal{C}}_{11}^\text{st}) + \delta^2$, $b_3=(1-p_\text{st}) \tilde{\mathcal{C}}_{11}^\text{st} + \delta^2$, and $b_4=(1-p_\text{st})(1-\tilde{\mathcal{C}}_{11}^\text{st})-\delta^2$. Finally, the entanglement negativity reads
\begin{align} \label{neganst}
	\mathcal{N}=\max(0,-\lambda_1),
\end{align}
where
\begin{align} \label{lambda1ver1}
	&\lambda_1=\frac{1}{2} \left[1-f_L-f_R+2f_L f_R \right. \\ \nonumber & \left. -\sqrt{1-2(f_L+f_R)+\Delta f^2+4p_\text{st}(f_L+f_R-p_\text{st})}\right],
\end{align}
with $\Delta f=f_L-f_R$.

Let us now rewrite the expression above in terms of the system parameters. For the Markovian dynamics, $p_\text{st}$ is given by a solution of the master equation~\cite{fichetti1998}
\begin{align}
	\Gamma_L(f_L-p_\text{st})+\Gamma_R(f_R-p_\text{st})=0,
\end{align}
which yields
\begin{align}
	p_\text{st}=\frac{\Gamma_L f_L +\Gamma_R f_R}{\Gamma_L+\Gamma_R}.
\end{align}
Using a parameterization of tunneling rates and chemical potentials defined in Sec.~\ref{sec:voltmod}, the eigenvalue $\lambda_1$ takes the form
\begin{align} \label{lambda1ver2}
&\lambda_1= \frac{1}{2\left[\cosh(\beta\bar{\mu})+\cosh \left(\frac{\beta V}{2} \right) \right]} \times \left \{ \cosh(\beta \bar{\mu}) \right. \\ \nonumber
& \left. -\sqrt{\cosh^2(\beta \bar{ \mu})+\frac{1}{2} \left[(1-a^2) \cosh(\beta V)+a^2-3 \right]} \right \}.
\end{align}

\subsubsection{Analysis of the results} \label{subsec:resan}
\begin{figure}
	\centering
	\includegraphics[width=0.9\linewidth]{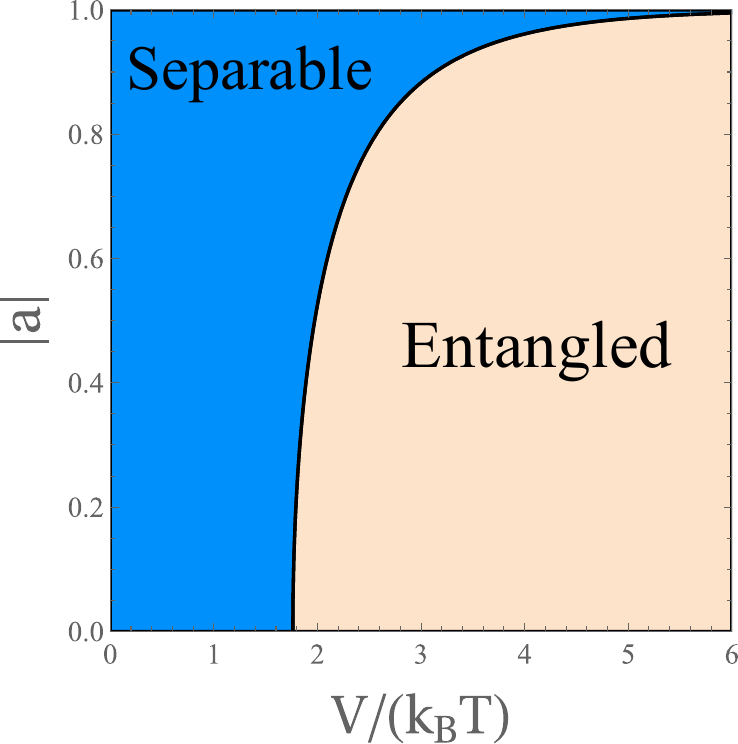}
	\caption{The entanglement phase diagram showing parameter regions in which the system-bath state is entangled or separable.}
	\label{fig:phasediagst}
\end{figure}
We now analyze consequences of Eqs.~\eqref{neganst} and~\eqref{lambda1ver2}. First, as implied by Fig.~\ref{fig:transient}~(b), the entanglement appears above a certain threshold voltage $V_\text{th}$. It can be found by solving $\lambda_1=0$. From Eq.~\eqref{lambda1ver2}, this is equivalent to solving the equation
\begin{align}
	(1-a^2) \cosh(\beta V_\text{th})+a^2-3=0,
\end{align}
which is independent of the average chemical potential $\bar{\mu}$. Thus, $\bar{\mu}$ determines only the magnitude of the system-bath entanglement, but not its presence. The solution reads
\begin{align}
	V_\text{th}=k_B T \text{arccosh} \left( \frac{3-a^2}{1-a^2} \right).
\end{align}
Equivalently, for a given voltage $V$, the entanglement is present for
\begin{align} \label{amax}
	|a| < \sqrt{\frac{\cosh(\beta V)-3}{\cosh(\beta V)-1}}.
\end{align}
Parameter regions in which the entanglement is present or absent are presented graphically in the entanglement phase diagram (Fig.~\ref{fig:phasediagst}). As one can note, the threshold voltage increases with the asymmetry coefficient $a$. In particular, it vanishes in the limit $|a| \rightarrow 1$, when the system is effectively coupled to a single bath. In fact, this regime is equivalent to the equilibrium case, where no entanglement is present in the weak-coupling regime. However, for large voltages $V \gtrapprox 5 k_B T$ the entanglement is present up to very large degrees of asymmetry, i.e., for
\begin{align}
	|a| \lessapprox 1-2e^{-\beta V},
\end{align}
where the right-hand side of inequality is very close to 1.

\begin{figure}
	\centering
	\includegraphics[width=0.9\linewidth]{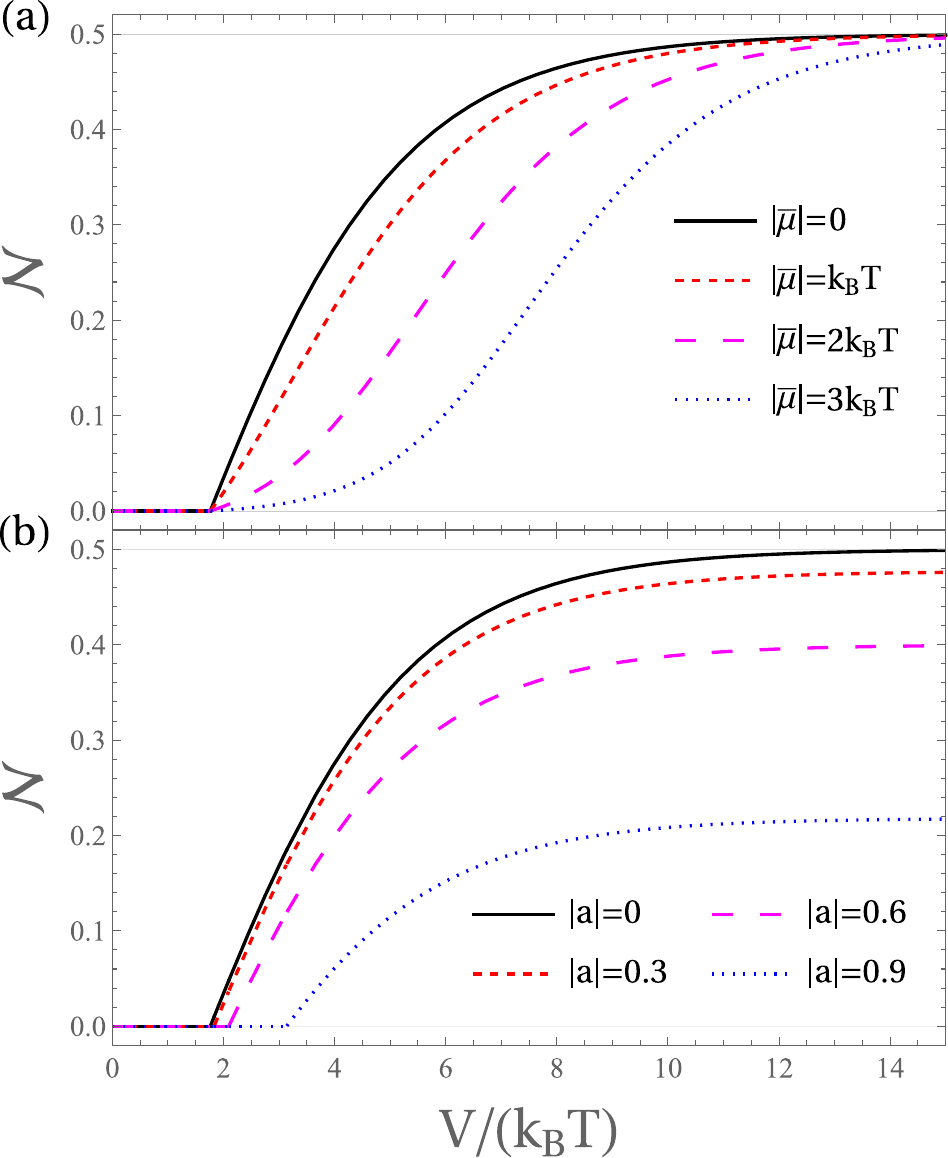}
	\caption{Entanglement negativity $\mathcal{N}$ as a function of voltage $V$ for different average chemical potentials $\bar{\mu}$ (a) and tunnel coupling asymmetries $a$ (b).}
	\label{fig:voltdep2}
\end{figure}

Let us now consider how entanglement is quantitatively affected by either the finite average chemical potential $\bar{\mu}$, which breaks the particle-hole symmetry [Fig.~\ref{fig:voltdep2}~(a)], or by the tunnel coupling asymmetry $a$ [Fig.~\ref{fig:voltdep2}~(b)]. First, as already noted, the average chemical potential affects the magnitude of the entanglement negativity for $|\bar{\mu}|$ comparable to $V$, but not the threshold voltage at which it appears. This somewhat resembles a similar previous result obtained for the equilibrium strong-coupling regime [Fig.~\ref{fig:eqgdep}]. However, the trend is reversed compared to that observed for the equilibrium case: The magnitude of entanglement decreases with the absolute value $\bar{\mu}$. A possible explanation may be that by increasing $|\bar{\mu}|$ one suppresses the particle flow between the baths, which is the source of the steady-state entanglement [as the particle current is proportional to $\Gamma(f_L-f_R)$, which is maximized at $\bar{\mu}=0$]. However, we note that breaking of the particle-hole symmetry no longer plays a role in the high-voltage regime, where $f_L \rightarrow 1$ and $f_R \rightarrow 0$ (independently of $\bar{\mu}$). Thus, for a high voltage, the entanglement negativity converges to the asymptotic value $\mathcal{N}=1/2$. 

In contrast, the asymmetry of tunnel couplings affects both the threshold voltage and the asymptotic value of the entanglement negativity in the high-voltage regime. Indeed, the latter value can be found analytically as
\begin{align}
	\lim_{V \rightarrow \infty} \mathcal{N}=\frac{\sqrt{1-a^2}}{2}.
\end{align}
 
\subsection{Steady state -- numerical results} \label{subsec:ststnum}

\begin{figure}
	\centering
	\includegraphics[width=0.9\linewidth]{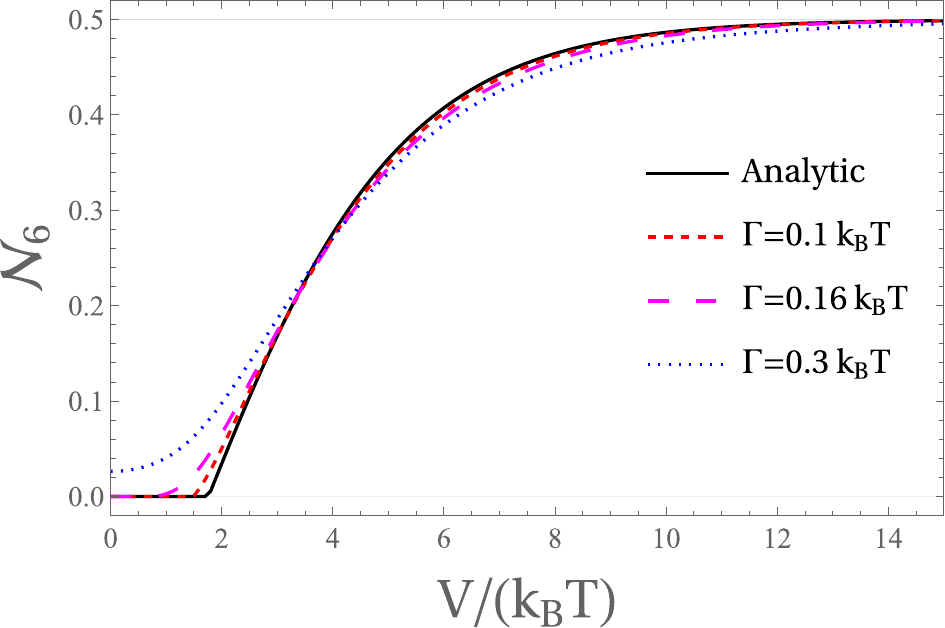}
	\caption{Entanglement negativity $\mathcal{N}_6$ as a function of voltage $V$ at a fixed time $t=10 \Gamma^{-1}$ for different coupling strengths $\Gamma$. Results for $p_0=0.5$, $a=0$, $\epsilon_0=\bar{\mu}=0$, $W=50 \Gamma$, and $K=300$.}
	\label{fig:gamma-voltdep}
\end{figure}
Let us now analyze the numerical results to establish a range of validity of the Markovian theory, as well as go beyond this regime. To this end, we analyze the partial negativity $\mathcal{N}_6$ at a fixed time $t=10\Gamma^{-1}$, which is much longer than the relaxation time. In Fig.~\ref{fig:gamma-voltdep}, we present the voltage dependence of the steady-state entanglement for different coupling strengths $\Gamma$, focusing on the highly symmetric case with $\bar{\mu}=\epsilon_0=0$ and $a=0$. As can be observed, for a weak coupling strength $\Gamma=0.1 k_B T$, the entanglement negativity agrees well with the Markovian theory. For an intermediate coupling $\Gamma=0.16 k_B T$, the threshold voltage is shifted to a much lower value. Finally, in the strong-coupling regime ($\Gamma=0.3 k_B T$) entanglement is also present in equilibrium ($V=0$). 

\begin{figure}
	\centering
	\includegraphics[width=0.9\linewidth]{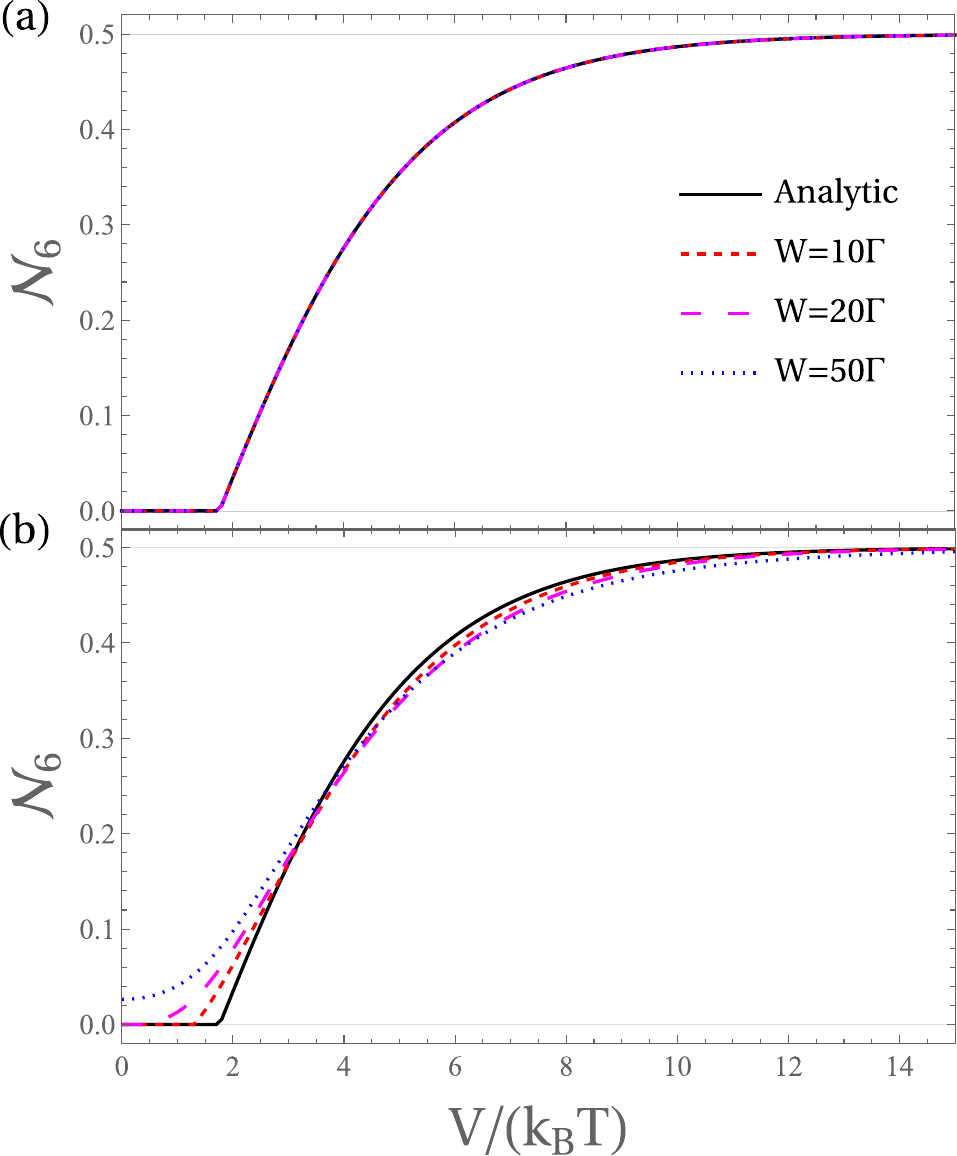}
	\caption{Entanglement negativity $\mathcal{N}_6$ as a function of voltage $V$ at a fixed time $t=10 \Gamma^{-1}$ for different bandwidths $W$ with (a) $\Gamma=0.01 k_B T$ and (b) $\Gamma=0.3 k_B T$. Other parameters as in Fig.~\ref{fig:gamma-voltdep}.}
	\label{fig:band-voltdep}
\end{figure}
\begin{figure}
	\centering
	\includegraphics[width=0.9\linewidth]{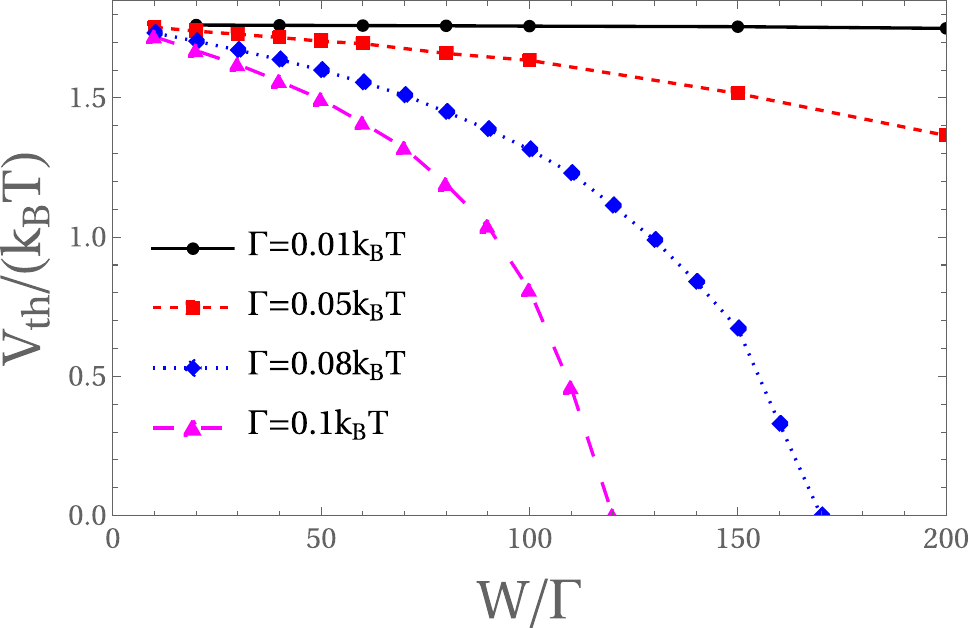}
	\caption{The threshold voltage $V_\text{th}$, at which the entanglement appears, as a function of the ratio $W/\Gamma$, for different coupling strengths $\Gamma$, evaluated for $M=8$. Other parameters as in Fig.~\ref{fig:gamma-voltdep}.}
	\label{fig:vth1}
\end{figure}
In Fig.~\ref{fig:band-voltdep}, as in the equilibrium case, we analyze the dependence of entanglement on the bandwidth, focusing on a range of bandwidths when the system reaches a steady state independent of the initial conditions (cf. Sec.~\ref{subsec:volttrans}). For a weak coupling $\Gamma=0.01 k_B T$, the entanglement almost does not depend on the bandwidth in the whole range of $W$ that we can simulate, and it agrees with the predictions of the analytic theory. For a stronger coupling ($\Gamma=0.3 k_B T$), by increasing the bandwidth we decrease the threshold voltage $V_\text{th}$ at which the entanglement appears, such that above a certain bandwidth entanglement is also present at equilibrium. The dependence of the threshold voltage on the bandwidth is further presented in Fig.~\ref{fig:vth1}. As shown, it monotonically decreases with the bandwidth. Furthermore, the critical bandwidth $W$, at which the threshold voltage goes to $0$, decreases as we increase the coupling strength.

\begin{figure}
	\centering
	\includegraphics[width=0.9\linewidth]{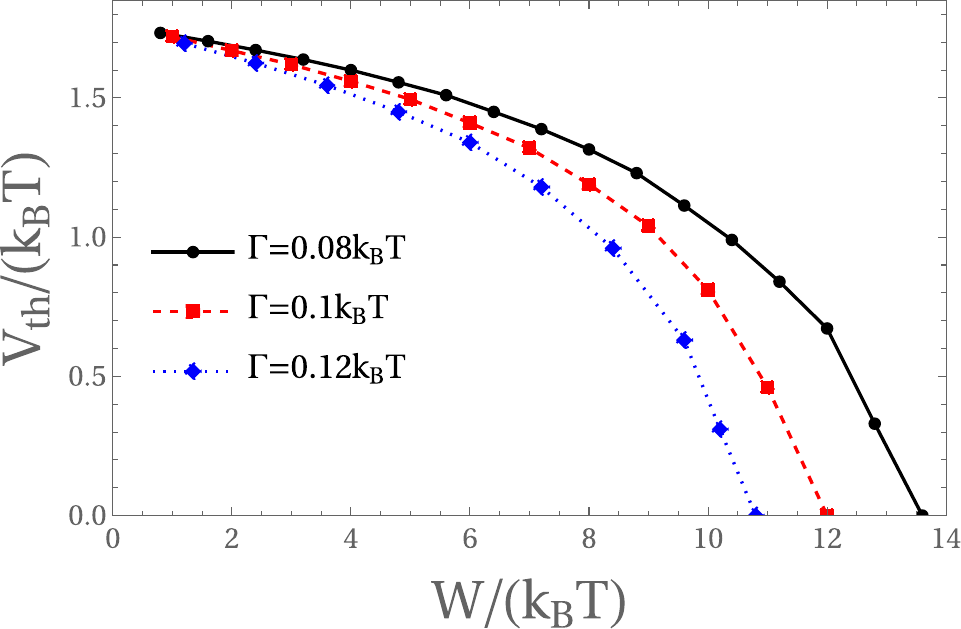}
	\caption{The threshold voltage $V_\text{th}$, at which the entanglement appears, as a function of the ratio $W/(k_B T)$, for different coupling strengths $\Gamma$, evaluated for $M=8$. Other parameters as in Fig.~\ref{fig:gamma-voltdep}.}
	\label{fig:vth2}
\end{figure}
We note that we related the bandwidth-dependence of entanglement to coupling to highly pure off-resonant levels in the bath. The high purity is the result of the thermal occupancy of the levels going to 0 or 1, which is determined (via the Fermi distribution) by the ratio of $\epsilon_k-\mu$ and the thermal energy $k_B T$. Therefore, it might be relevant to plot the dependence of the threshold voltage as a function of the ratio $W/(k_B T)$ rather than $W/\Gamma$. This is done in Fig.~\ref{fig:vth2}. Here one can still observe that the critical ratio $W/(k_B T)$, at which the threshold voltage goes to 0, increases as $\Gamma$ decreases. This raises the question of what happens in the joint limit $\Gamma \rightarrow 0$ and $W \rightarrow \infty$, which is often used to define a weak-coupling Markovian regime~\cite{schaller2014}. Is the entanglement present only above a certain finite threshold voltage? Is it present at any finite voltage? Or is it present even at equilibrium? We cannot answer this question conclusively using our simulations.

\section{Conclusions} \label{sec:concl}
In this paper we investigated the behavior of entanglement between a single fermionic energy level and a fermionic bath in different thermodynamic regimes. We first considered entanglement in the global equilibrium state of the system and the bath. For the grand canonical state (with fluctuating energy and particle number) entanglement appears for a finite coupling strength of the order of $k_B T$. Quite notably, this threshold coupling strength can be decreased by increasing the bath bandwidth. Interestingly, this implies that the presence of entanglement may be affected by the spectral density of the bath at energies strongly off-resonant with the system. We relate this effect to the correlation with highly pure levels in the bath (with thermal occupancy close to 0 or 1), which, even though quantitatively weak, tends to be genuinely quantum rather than classical. There are even hints that entanglement may be present for any finite coupling strength in the infinite bandwidth limit. However, we cannot confirm this conclusively using our simulations. The magnitude of entanglement (but not its presence) further depends on the degree of particle-hole symmetry breaking (i.e., relative position of the system energy and the chemical potential).

Furthermore, our study revealed the dependence of entanglement on the considered statistical ensemble, which does not affect the reduced state of the system (according to the principle of ensemble equivalence). In contrast to the case discussed above, for the canonical ensemble with a fixed particle number, entanglement appears for arbitrarily weak system-bath couplings. As follows from the theory presented in Ref.~\cite{ma2022}, this is a result of coexistence of quantum coherence in the Fock basis and charge conservation.

We then investigated the behavior of entanglement during relaxation of the impurity initialized in an out-of-equilibrium state and attached to a single bath. First of all, we derived an analytic theory describing entanglement in the weak system-bath coupling regime (for sufficiently small bandwidths). Its validity is further confirmed by numerical simulations, which also enables us to go beyond the weak-coupling regime. Our results show that a transient system-bath entanglement can be generated even in the regime where the system dynamic can be well described by an effectively classical Markovian master equation for the system occupancy. This shows that, for fermionic systems, the validity of Born-Markov approximation, and the possibility of an effectively classical description of the reduced dynamics, do not preclude the existence of system-bath entanglement. 

While in the weak-coupling case the entanglement tends to ultimately vanish (as the system-bath state tends to asymptotically factorize~\cite{cusumano2018}), for a stronger coupling it tends to saturate at a finite value, consistent with the value for the global thermal state of the total system-bath Hamiltonian. This conclusion holds provided that transient dynamics leads to thermalization of the system. This may be suppressed, e.g., by strongly non-Markovian effects related to the presence of the bound states. In such a case, entanglement may be generated and preserved at long times even when there is no thermal entanglement for the same system parameters. Furthermore, in contrast to Ref.~\cite{bernardo2021}, we found no direct link between entanglement and the system-bath interaction energy. Indeed, entanglement can be observed even in the weak-coupling regime, where the interaction energy vanishes. This contradicts the conclusion of Ref.~\cite{bernardo2021} that the interaction energy is responsible for the entanglement generation.

Finally, we covered the case of a voltage-driven junction consisting of an impurity attached to two reservoirs with different chemical potentials. Using a derived analytic theory, we showed that the system-bath entanglement is generated for an arbitrarily weak coupling to the reservoirs at a certain threshold voltage, which increases with the asymmetry of the tunnel couplings. The entanglement magnitude is reduced also by the deviation from the particle-hole symmetry, which does not affect the threshold voltage. For a stronger coupling, analogously to the equilibrium case, the threshold voltage is further reduced by increasing the bath bandwidth.

Overall, our results suggest that the system-bath entanglement is quite ubiquitous in fermionic systems, as it can appear even under relatively mild conditions (such as a weak system-bath coupling for a large bandwidth in the equilibrium case, or moderate voltages in the nonequilibrium steady state). In particular, it can be present even when many aspects of the system behavior are effectively classical, e.g., during the Markovian relaxation process. Therefore, one must be careful when associating the presence of entanglement with nontrivial quantum phenomena, such as strong electronic correlations or strong-coupling thermodynamic effects.

Let us now consider potential future research directions motivated by our results. First, while our study focused on entanglement with the whole bath, it might be interesting to investigate the spatial extension of entanglement (motivated by previous research on the Kondo cloud~\cite{bayat2010, lee2015, wagner2018, shim2023}). In particular, we expect that the grand canonical and canonical ensembles should predict the same entanglement with a neighboring region of the impurity (as the reduced state of this region is the same for both ensembles), while deviation should appear at larger distances. Second, an obvious research direction is to explore the role of interelectron interactions (e.g., the Kondo effect) in generating and preserving the system-bath entanglement out of equilibrium (in both transient and steady-state regimes), or interplay between interactions and the bath bandwidth. This may be done by the numerical renormalization group approach proposed in Ref.~\cite{shim2018}, which can hopefully be generalized to the nonequilibrium case~\cite{anders2005, anders2006}. Finally, it might be interesting to study multilevel systems. This will allow one to explore the role of initial intrasystem coherence, or level degeneracy and quantum interference, previously studied in the context of system-bath mutual information~\cite{dey2020}.

\acknowledgments
This research was supported by the FQXi foundation Project No.~FQXi-IAF19-05-52 ``Colloids and superconducting quantum circuits''.

\appendix

\section{Partial transposition} \label{sec:transp}
To define the partial transposition, let us consider a generic bipartite quantum system $AB$ with the basis states of $A$ and $B$ denoted as $|i \rangle$, $|j \rangle$ and $|k \rangle$, $|l \rangle$, respectively. Then the density matrix of the bipartite system $\rho_{AB}$, with subsystems having the Hilbert space dimensions $\text{dim}_A=P$ and $\text{dim}_B=R$, can be written as a $P \times P$ block matrix 
\begin{align}
	\rho_{AB} = \begin{pmatrix} 
		\gamma_{11} & \gamma_{12} & \dots \\
		\vdots & \ddots & \\
		\gamma_{P1} &        & \gamma_{PP}
	\end{pmatrix},
\end{align}
where blocks $\gamma_{ij}$ are square matrices of size $R \times R$ defined as $(\gamma_{ij})_{kl}={\text{Tr} [\rho_{AB} (|i \rangle \langle j| \otimes |k \rangle \langle l|)]}$. Then, the partial transpose of the state of the subsystem $B$ is defined as~\cite{peres1996, horodecki1996}
\begin{align}
	\rho_{AB}^{T_B} = \begin{pmatrix} 
		\gamma_{11}^T & \gamma_{12}^T & \dots \\
		\vdots & \ddots & \\
		\gamma_{P1}^T &        & \gamma_{PP}^T
	\end{pmatrix}.
\end{align}

\section{Householder tridiagonalization} \label{sec:hous}
Here we discuss how one can tridiagonalize the Hermitian correlation matrix $\mathcal{C}$ by means of the Householder transformation~\cite{householder1958}. To this goal, we apply a simple algorithm presented in Ref.~\cite{ozaki}; we rewrite it here for the sake of completeness of the paper.

Let us first write the correlation matrix in the block diagonal form
\begin{align}
	\mathcal{C} = \begin{pmatrix} 
		\mathcal{C}_{00} & \mathbf{b}^\dagger \\
		\mathbf{b} & \mathcal{C}_B
	\end{pmatrix},
\end{align}
where $\mathcal{C}_B$ is the reduced correlation matrix of the bath (corresponding to modes $i=1,\ldots,K$) and $\mathbf{b}$ is the column vector
\begin{align}
	\mathbf{b}=(\mathcal{C}_{01},\ldots,\mathcal{C}_{0K})^\dagger.
\end{align}
We then define the column vectors of size $K$.
\begin{align}
	\mathbf{s} &=(\mathbf{b}^\dagger \mathbf{b},0,\ldots,0)^T, \\
	\mathbf{v} &=\frac{\mathbf{b}-\mathbf{s}}{\sqrt{(\mathbf{b}-\mathbf{s})^\dagger (\mathbf{b}-\mathbf{s})}},
\end{align}
and the parameters
\begin{align}
	\alpha_r &=\frac{1}{2} \left(2\mathbf{s}^\dagger \mathbf{s}-\mathbf{b}^\dagger \mathbf{s}-\mathbf{s}^\dagger \mathbf{b}\right), \\
	\alpha_i&=-\text{Im} (\mathbf{b}^\dagger \mathbf{s}), \\
	\alpha&=\frac{2\alpha_r}{\alpha_r^2+\alpha_i^2} (\alpha_r+i\alpha_i).
\end{align}
The correlation matrix can then be unitarily transformed to a form
\begin{align}
	\tilde{\mathcal{C}}=\begin{pmatrix} 
		\mathcal{C}_{00} & \mathbf{s}^\dagger \\
		\mathbf{s} & 	\tilde{\mathcal{C}}_B
	\end{pmatrix},
\end{align}
where 
\begin{align}
	\tilde{\mathcal{C}}_B&=\mathcal{Q}^\dagger \mathcal{C}_B \mathcal{Q}, \\
	\mathcal{Q}&=\mathds{1}_K-\alpha \mathbf{v} \mathbf{v}^\dagger,
\end{align}
and $\mathds{1}_K$ is the identity matrix of size $K$. In this form, all off-diagonal elements $\tilde{\mathcal{C}}_{0i}=\tilde{\mathcal{C}}_{i0}^*$ apart from $\tilde{\mathcal{C}}_{01}=\tilde{\mathcal{C}}_{10}=\mathbf{b}^\dagger \mathbf{b}$ vanish, which is a first step of tridiagonalization. Complete tridiagonalization can be performed iteratively applying the same procedure to the matrix $\tilde{\mathcal{C}}_B$, etc.

\section{System-bath mutual information} \label{sec:mutinf}
While our paper focuses on the entanglement negativity, which measures genuine quantum correlations, let us here apply our analytic theory from Sec.~\ref{subsec:ander} for the study of system-bath mutual information, previously investigated numerically in Ref.~\cite{ptaszynski2022}. This quantity is defined as
\begin{align}
	I_{SB}=S_S+S_B-S_{SB},
\end{align}
where $S_\alpha=-\text{Tr} (\rho_\alpha \ln \rho_\alpha)$ ($\alpha \in \{S,B,SB\}$) is the von Neumann entropy. The mutual information measures the total system-bath correlations (both classical and quantum). For the model analyzed, it can be evaluated using the density matrix given by Eq.~\eqref{denmatan} as
\begin{align} \label{mutinfan}
	I_{SB}=h(p_t)+h(f-p_t+p_0)-h(p_0)-h(f),
\end{align} 
where $h(x)=-x \ln x -(1-x) \ln (1-x)$ is the binary entropy. 

\begin{figure}
	\centering
	\includegraphics[width=0.9\linewidth]{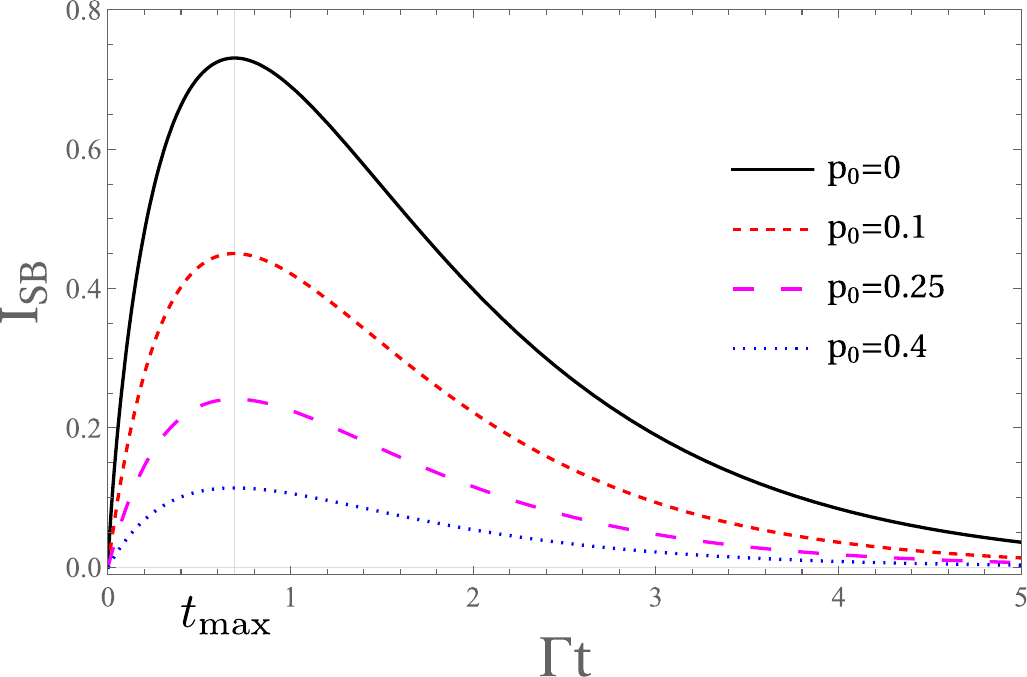}
	\caption{System-bath mutual information as a function of time for different initial system occupancies $p_0$. Results for $\mu=\epsilon_0+k_B T$.}
	\label{fig:mutinf}
\end{figure}

We further focus on the case when the system occupancy follows a Markovian relaxation dynamics given by Eq.~\eqref{markev}. The evolution of the system-bath mutual information for different initial conditions is presented in Fig.~\ref{fig:mutinf}. As one can observe, its behavior is non-monotonic: the mutual information is first generated, reaches a maximum value, and later asymptotically decays to 0. The same behavior was demonstrated numerically in Ref.~\cite{ptaszynski2022}, where the long-time decay of mutual information has been explained as a result of its reconversion into the correlations within the bath. An analogous dynamics of $I_{SB}$ has been also observed for noninteracting bosonic systems~\cite{pucci2013, einsiedler2020, colla2021}.

As the behavior of the system-bath mutual information has been thoroughly investigated numerically, it does not need to be analyzed in detail. However, the analytic theory provides a certain qualitative insight that goes beyond the numerics. First, analogously to the entanglement negativity, the mutual information has a universal maximum at $t_\text{max}=\Gamma^{-1} \ln 2$ and obeys a symmetry relation
\begin{align}
	I_{SB}(t_1) = I_{SB}(t_2) \quad \text{for} \quad e^{-\Gamma t_1}=1-e^{-\Gamma t_2}.
\end{align}
Furthermore, the analytic theory enables us to analyze the long-time asymptotic behavior of $I_{SB}$. Expanding Eq.~\eqref{mutinfan} to the lowest order of $p_t-f=(p_0-f)e^{-\Gamma t}$ one finds
\begin{align}
	I_{SB} = \ln \frac{p_0(1-f)}{(1-p_0)f} \times e^{-\Gamma t} + \mathcal{O}(e^{-2\Gamma t}).
\end{align}
Thus, for long times the system-bath mutual information undergoes an exponential decay with a relaxation rate $\Gamma$. However, this expansion is not applicable for an initial pure state ($p_0=0$ or $p_0=1$), when the pre-exponential factor diverges. In this case, the mutual information at long times follows an exponential decay slowed down by a linearly increasing prefactor:
\begin{align}
	I_{SB} \approx \Gamma t e^{-\Gamma t} \quad \text{for} \quad t \gg \Gamma^{-1}.
\end{align}
Indeed, as shown in Fig.~\ref{fig:mutinf}, the decay of mutual information is slower for an initial pure state than for initial mixed states.

\end{document}